\documentclass[pre,amsmath,superscriptaddress,floatfix]{revtex4}

\usepackage{rotating}
\usepackage{color}
\usepackage{amsmath,amstext,amsgen,amsfonts}
\usepackage{epsfig}
\def \kv{\mathbf{k}}
\def \h{\"}

\begin{document}

%%%%%%%%%%%%%%%%%%

\title{Dynamical Systems on Networks: A Tutorial}

\vspace{-.2in}
\author{Mason A. Porter}
% (\emph{porterm@maths.ox.ac.uk})}
\affiliation{Oxford Centre for Industrial and Applied Mathematics,
Mathematical Institute, University of Oxford, Oxford, OX2 6GG, UK}
\affiliation{CABDyN Complexity Centre, University of Oxford, Oxford, OX1 1HP, UK}
\author{James P. Gleeson}
\affiliation{MACSI, Department of Mathematics \& Statistics, University of Limerick, Ireland}

%%%%%%%%%%%%

\begin{abstract}

We give a tutorial for the study of dynamical systems on networks. We focus especially on ``simple" situations that are tractable analytically, because they can be very insightful and provide useful springboards for the study of more complicated scenarios. We briefly motivate why examining dynamical systems on networks is interesting and important, and we then give several fascinating examples and discuss some theoretical results.  We also briefly discuss dynamical systems on dynamical (i.e., time-dependent) networks, overview software implementations, and give an outlook on the field.

\end{abstract}

%%%%%%%

\maketitle

%%%%%%%%%%%%%%

\section{Preamble}

The purpose of the present paper is to give a tutorial for studying dynamical systems on networks. We explicitly focus on ``simple" situations that are tractable analytically, though it is also valuable to examine more complicated situations (and insights from simple scenarios can help guide such investigations).  Our article is intended for people who seek to study dynamical systems on networks but who might not have any prior experience at it.  Accordingly, we hope that reading it will convey why it is both interesting and useful to study dynamical systems on networks, how one can go about doing so, the potential pitfalls that can arise in such studies, the current research frontier in the field, and important open problems.

We also hope that reading our article will provide helpful pointers to interesting papers to read, and that it will facilitate your ability to critically read and evaluate papers that concern dynamical systems on networks.  Our article is \emph{not} a review (or anything close to one), and in particular we are citing only a small subset of the existing scholarship in this area.  New articles on dynamical systems on networks are published or posted on a preprint server very frequently, so we couldn't possibly cite all of the potentially relevant articles comprehensively even if we tried. See Sec.~\ref{sec:OtherResources} for books, review articles, surveys, and tutorials on various related topics.

%%%%%%%%%%%

\section{Introduction: How Does Nontrivial Network Connectivity Affect Dynamical Processes on Networks?}

When studying a dynamical process, one is concerned with its behavior as a function of time, space, and its parameters. There are numerous studies that examine how many people are infected by a biological contagion and whether it persists from one season to another, whether and to what extent interacting oscillators synchronize, whether a meme on the internet becomes viral or not, and more.  These studies all have something in common: the dynamics are occurring on a set of entities (\emph{nodes} in a network) that are connected to each other via \emph{edges} in some nontrivial way.  This leads to the natural question of how such underlying nontrivial connectivity affects dynamical processes.  This is one of the most important questions in network science \cite{newman2010}, and this is the core question that we consider in the present article.

Traditional studies of continuous dynamical systems are concerned with qualitative methods to study coupled ordinary differential equations (ODEs) \cite{stro1994,guckholmes} and/or partial differential equations (PDEs) \cite{cross,predrag}, and traditional studies of discrete dynamical systems take analogous approaches with maps \cite{stro1994,guckholmes}.\footnote{Of course, nothing is stopping us from placing more complicated dynamical processes---which can be governed by stochastic differential equations, delay differential equations, or something else---on a network's nodes.} If the state of each node in a network is governed by its own ODE (or PDE or map), then studying a dynamical process on a network entails examining a (typically large) system of coupled ODEs (or PDEs or maps). The change in state of a node depends not only on its own current state but also on those of its neighbors, and the network encodes which nodes interact with each other and how strongly they interact.

An area of particular interest (because of tractability and seeming simplicity) is binary-state dynamics on nodes, whose states depend on those of their neighbors and which often have stochastic update rules. (Dynamical processes with more than two states are obviously also interesting.)
%, but binary situations are often convenient to study because they are tractable.)
Examples include simple models of disease spread, where each node is considered to be in either a healthy (\emph{susceptible}) state or an unhealthy (\emph{infected}) state, and infections are transmitted probabilistically along the edges of a network.  One can apply approximation methods, such as mean-field approaches, to obtain (relatively) low-dimensional descriptions of the global behavior of the system---e.g., to predict the expected number of infected people in a network at a given time---and these methods can yield ODE systems that are amenable to analysis via standard approaches from the theory of dynamical systems.

Moreover, it is true not only that network structure can affect dynamical processes on a network, but also that dynamical processes can affect the dynamics of the network itself. For example, when a child gets the flu, he/she might not go to school for a couple of days, and this temporary change in human activity affects which social contacts take place, which can in turn affect the dynamics of disease propagation.  We will briefly discuss the interactions of dynamics on networks with dynamics of networks (these are sometimes called ``adaptive networks" \cite{thilo-adaptive,thilo-adapt2}) in this article, but we will mostly assume static network connectivity so that we can focus on the question of how network structure affects dynamical processes that run on top of a network. Whether this is reasonable for a given situation depends on the relative timescales of the dynamics on the network and the dynamics of the network.

%%%%%

%\subsection{Organization of This Article}

The remainder of this article is organized as follows. Before delving into dynamics, we start by recalling a few basic concepts in Sec.~\ref{basic}.  In Sec.~\ref{examples}, we discuss several examples of dynamical systems on networks.  In Sec.~\ref{general}, we give various theoretical considerations for general dynamical systems on networks as well as for several systems on which we focus. We overview software implementations in Sec.~\ref{numericalmethods}.  In Sec.~\ref{temporal}, we briefly examine dynamical systems on dynamical (i.e., time-dependent) networks, and we recommend several resources for further reading in Sec.~\ref{sec:OtherResources}.   Finally, we conclude and discuss both open problems and current research efforts in Sec.~\ref{conclude}.

%%%%%%%%

\section{A Few Basic Concepts}\label{basic}

For simplicity, we will frame our discussions in terms of unweighted, undirected networks. When such a network is static, it can be represented using a symmetric adjacency matrix ${\bf A} = {\bf A}^T$ with elements $A_{ij} = A_{ji}$ that are equal to $1$ if nodes $i$ and $j$ are connected (or, more properly, ``adjacent'') and $0$ if they are not.  We will also assume that $A_{ii} = 0$ for all $i$, so that none of our networks include self-edges. We denote the total number of nodes in a network (i.e., its ``size'') by $N$. The \emph{degree} $k_i$ of node $i$ is the number of edges that are connected to it. For a large network, it is common to examine the distribution of degrees over all of its nodes.  This \emph{degree distribution} $P_k$ is defined as the probability that a node---chosen uniformly at random from the set of all nodes---has degree $k$. The \emph{mean degree} $z$ is the mean number of edges per node and is given by $z=\sum_k k P_k$. For example, classical Erd\H{o}s-R\'enyi (ER) random graphs have a Poisson degree distribution: $P_k=\frac{z^k e^{-z}}{k!}$. However, many real-world networks have right-skewed (i.e., \emph{heavy-tailed}) degree distributions \cite{Clauset2009}, so the mean degree $z$ only provides minimal information about structure of a network. The most popular type of heavy-tailed distribution is a \emph{power law} \cite{Stumpf2012}, for which $P_k\sim k^{-\gamma}$ as $k\to\infty$ (where the parameter $\gamma$ is called the ``power" or ``exponent").  Networks with a power-law degree distribution are often called by the name of ``scale-free networks" (though such networks can still have scales in them, so the monicker is misleading), and many generative mechanisms---such as de Solla Price's model \cite{price76} and the Barab\'asi-Albert (BA) model \cite{Barabasi99}---have been developed to produce networks with power-law degree distributions.

The effects of network structure on dynamics are often studied using a random-graph ensemble known as the \emph{configuration model} \cite{bollobas-random,newman2010}. In this ensemble, one specifies the degree distribution $P_k$, but the network \emph{stubs} (i.e., ends of edges) are then connected to each other uniformly at random. In the limit of infinite network size, one expects a network drawn from a configuration-model ensemble to have vanishingly small \emph{degree-degree correlations} and \emph{local clustering} \footnote{Strictly speaking, one also needs to ensure appropriate conditions on the moments $P_k$ as $N \rightarrow \infty$. For example, one could demand that the mean degree and variance increase sufficiently slowly in the infinite-size limit. It is believed that such conditions tend to hold in ensembles of real-world networks, though it tends to be difficult to procure ensembles of differently-sized networks from real data.  (A notable exception is the Facebook networks from \cite{Traud2012}.)}. Degree-degree correlation measures the (Pearson) correlation between the degrees of nodes at each end of a randomly-chosen edge of a network.  (The edge is chosen uniformly at random from the set of edges.)  Degree-degree correlation can be significant, for example, if high-degree nodes are connected preferentially to other high-degree nodes.  This would be true in a social network if popular people tend to be friends with other popular people, and one would describe the network as homophilous by degree.  By contrast, a network for which high-degree nodes are connected preferentially to low-degree nodes is heterophilous by degree. The simplest type of local clustering arises as a result of a preponderance of triangle motifs in a network. (More complicated types of clustering---which need not be local---include motifs with more than three nodes, community structure, and core-periphery structure \cite{Porter2009,newman2010,cp-review}.) Triangles are common, for example, in social networks, so the lack of local clustering in configuration-model networks (in the thermodynamic limit) is an important respect in which their structure differs significantly from that in most real networks. Investigations of dynamical systems on networks with different types of clustering is a focus of current research \cite{Adamcascades,Piikk,Miller15}.

%%%%%

\section{Examples of Dynamical Systems} \label{examples}

A huge number of dynamical systems have been studied in numerous disciplines and from multiple perspectives, and an increasingly large number of these systems have also been examined on networks.  In this section, we present examples of some of the most prominent dynamical systems that have been studied on networks. We purposely focus on ``simple" situations that are tractable analytically, though studying more complicated systems---typically through direct numerical simulations---is also worthwhile. 

Many of the dynamical processes that we consider can of course be studied in much more complicated situations (including in directed networks, weighted networks, temporal networks \cite{thilo-adaptive}, and multilayer \cite{mikko-review,bocca-review} networks), and many interesting new phenomena occur in these situations.  In this article, however, we want to keep network structures as simple as possible.  We will discuss ways in which network structure has a nontrivial impact on dynamical processes, but we will have minimal discussion of the aforementioned complications\footnote{The complication that we will discuss the most are time-dependent network structures (i.e., temporal networks), because consideration of the time scales of dynamical processes on networks versus those of the dynamics of the networks themselves is a crucial modeling issue.}.
When we place a dynamical process on a network, we will sometimes refer to that network as a ``substrate''.

In this section, we will discuss examples of both discrete-state and continuous-state dynamical systems. For the former, it is important to consider whether to update node states synchronous or asychronously, so we also include an interlude that is devoted to this issue.

%%%%%%

\subsection{Percolation}\label{percolation}

\emph{Percolation} theory is the study of qualitative changes in connectivity in systems (especially large ones) as its components are occupied or removed \cite{sethnabook}.  Percolation transitions provide an archetype for continuous transitions, and there has been a lot of work on percolation problems (especially on lattices but increasingly on more general networks) from both physical and mathematical perspectives \cite{newman2010,kesten-whatis}.

\subsubsection{Site Percolation}

The simplest type of percolation problem is \emph{site percolation} (i.e., \emph{node percolation}) \cite{newman2010}. Consider a network, and let each of its nodes be either occupied or unoccupied. One can construe occupied nodes as operational nodes in a network, whereas unoccupied nodes are nonfunctional. We might pick nodes uniformly at random and state that they are unoccupied (i.e., are effectively removed from the network) with uniform, independent probability $q = 1 - p \in [0,1]$; this is a so-called ``random attack" with \emph{occupation probability} $p$. Alternatively, we could be devious and perform some kind of ``targeted attack" in which we remove (i.e., set as unoccupied) some fraction of nodes preferentially by degree (which is, by far, the most usual case considered), geodesic node betweenness centrality (a measure of how often a node occurs on short paths), location in the network, or some other network diagnostic.  In the limit as the number of nodes $N \rightarrow \infty$ in one of these processes, what fraction $q_c$ of the nodes needs to be removed so that we no longer have a connected component---called a \emph{giant connected component (GCC)}---of occupied nodes?  A \emph{percolation transition} occurs at the critical occupation probability $p_c = 1 - q_c$, which is the point of appearance/disappearance of a GCC, which is a connected network component that scales in linear proportion to $N$ as $N \rightarrow \infty$.  (Such scaling is called ``extensive" in the language of statistical mechanics \cite{sethnabook}.)

%%%%%

\subsubsection{Bond Percolation}

In \emph{bond percolation} (i.e., \emph{edge percolation}), one tracks occupied edges instead of occupied nodes. Edges are labelled as occupied with probability $p$ (uniformly at random), which is called the \emph{bond occupation probability}. As with site percolation, the primary question of interest is the existence and size of a GCC, where connections occur only using occupied edges. If $p$ is below a critical value $p_c$, then too few edges are occupied to globally connect the network, and a GCC does not exist. However, above the threshold $p_c$, the number of nodes in the GCC is a finite fraction of $N$ as $N\to\infty$.

%%%%

\subsubsection{$K$-Core Percolation}

The \emph{$K$-core} of an unweighted, undirected network is the maximal subset of nodes such that each node is connected to at least $K$ other nodes \cite{seidman83}.  It is computationally fast to determine $K$-cores, and they are insightful for many situations \cite{Dorogovtsev06,cp-review}. In particular, every unweighted, undirected network has a \emph{$K$-core decomposition}.  Define a network's \emph{$K$-shell} as the set of all nodes that belong to the $K$-core but not to the $(K+1)$-core.  The network's $K$-core is given by the union of all $c$-shells for $c\geq{K}$, and the network's $K$-core decomposition is the set of all of its $c$-shells.

One can examine the $K$-core of a network as the limit of a dynamical pruning process. Starting with the initial network, delete all nodes with fewer than $K$ neighbors. After this pruning, the degree of some of the remaining nodes will have been reduced, so repeat the pruning step to delete nodes that now have fewer than $K$ remaining neighbors. Iterating this process until no further pruning is possible leaves the $K$-core of the network.

%%%%%

\subsubsection{``Explosive" Percolation} \label{kaboom}

%Fizz...  [I doubt there is a good place for me to put that]

A few years ago, Ref.~\cite{raissa-kaboom} suggested the possibility of an ``explosive" percolation process in which the transition (as a function of a parameter analogous to the bond occupation probability $p$) from a disconnected network to a network with a GCC is very steep and could perhaps even be discontinuous.  It has now been demonstrated that these family of ``explosive" processes, which are often called \emph{Achlioptos processes}, are in fact continuous \cite{oliver-science,oliver-aap,dorog2010}, but the very steep nature of the transitions that they exhibit has nevertheless fascinated many scholars.

Let's consider the simplest type of Achlioptas process. Start with $N$ isolated nodes and add undirected, unweighted edges one at a time.  Choose two possible edges uniformly (and independently) at random from the set of $N(N-1)/2$ possible edges between a pair of distinct nodes.  One then adds one of these edges, making a choice based on a systematic rule that affects the speed of development of a GCC (as compared to the analogous process in which one only picks a single edge by a random process).  One choice that yields ``explosive" percolation is to use the so-called ``product rule", in which one always retains the edge that minimizes the products of the sizes of the components that it joins (with an arbitrary choice when there is a tie).

Investigation of Achlioptas processes on different types of substrate networks and with different choices of rules---especially with the aim of developing rules that do not use global information about a graph---is an active area of research.

%%%%%%%

\subsubsection{Other Types of Percolation}

There are numerous other types of percolation, and it's worth bringing up a few more of them explicitly. \emph{Bootstrap} percolation is an infection-like process in which nodes become infected if sufficiently many of their neighbors are infected \cite{bootstrap,chalupa1979,adler1991}. It is related to the Centola-Macy threshold model for social contagions that we will discuss in Sec.~\ref{thresh}.  In \emph{limited path percolation}, one construes ``connectivity" as implying that a sufficiently short path still exists after some network components have been removed \cite{eduardo2007}.  (In this example, one can imagine trying to navigate a city but that some streets are blocked.) The percolation of $K$-cliques (i.e., completely connected subgraphs of $K$ nodes) has been used to study the algorithmic detection of dense sets of nodes known as ``communities" \cite{palla2005}. Various percolation processes have also been studied in several different types of multilayer networks (e.g., multiplex networks and interdependent networks) \cite{mikko-review,bocca-review}.

%%%

\subsection{Biological Contagions}\label{simplecontagions}

One of the standard ways to study biological contagions is through what are traditionally called \emph{compartmental models} (though such a term  could be used aptly to describe a much broader set of models), in which the compartments describe a state (e.g., ``susceptible", ``infected", or ``recovered") and there are parameters that represent transition rates for changing states \cite{ccc}. The simplest compartmental models are applicable to ``well-mixed'' populations, in which each individual can meet every other individual, and there each type of state change has a single associated probability. They can be described using ODEs if one is considering continuous-time state transitions or maps if one is considering discrete-time state transitions.  One can add complications by incorporating spatial considerations in the form of diffusion (via a Laplacian operator) or by constructing a metapopulation model, in which different populations (``patches") can have different fractions of entities in different states of an epidemic (or have other differences, such as different transition rates between their component states).  In a sense, a metapopulation model provides a simple way of incorporating network information, but one is really thinking of each node as representing some subpopulation of the full population rather than as an individual entity.  This distinction is important when one wishes to consider a metapopulation model on a network \cite{colizza2007-meta}.

The above frameworks assume that one is examining a well-mixed situation, but it is much more realistic in modern society to consider a network of contacts among agents \cite{rom-review2014,barratbook,newman2010,mollison1977,piet2014,marvel2013,durrettpnas2010}.  To do this, one places a compartmental model on a network, so that each node can be in one of several epidemic states (e.g., ``susceptible" or ``infected"), and the nodes have update rules that govern how the states change.  As we discuss in detail in Sec.~\ref{async}, the updates can occur either \emph{synchronously} or \emph{asynchronously}.  In synchronous updating, we consider discrete time, and all nodes are updated at once.  By contrast, in asynchronous updating, some small fraction of nodes---often just one node---are randomly chosen for update in each time step $dt$, or the updating algorithm can be event-driven.

Models of biological contagions are often called  ``simple contagions" because of the mechanism of an infection passing directly from one entity to another.  This is a reasonable toy model of some biological contagions, though of course real life can be significantly more complicated \cite{Hethcote00}. One can also examine biological contagions on more complicated types of networks (such as temporal \cite{holme13} and multilayer networks \cite{mikko-review,salehi2014}).

%%%%%%

\subsubsection{Susceptible-Infected (SI) Model}\label{SImodel}

The simplest type of biological epidemic has two states---\emph{susceptible} and \emph{infected}---where healthy nodes are considered ``susceptible" (and are in the ``S'' compartment) because they are not currently infected but can become infected, and ``infected" nodes (in the ``I'' compartment) permanently remain in that state.  This yields the \emph{susceptible-infected (SI) model}.

One can define the detailed dynamics of SI models in several different ways. The most common is to consider a stochastic process in which infection is ``transmitted'' from an infected node to a susceptible neighbor at a rate $\lambda$. This is a ``hazard rate", and the probability of a transmission event occurring on any chosen edge that connects an infected node to a susceptible node in an infinitesimal time interval $dt$ is $\lambda\, dt$. Suppose that we consider a susceptible node that has $m$ infected neighbors. The probability of this node becoming infected within the time interval $dt$ is then
\begin{equation}
	1-(1-\lambda \, dt)^m \to \lambda \, m \, dt \quad \mathrm{as} \quad  dt\to 0\,. \label{eqn1}
\end{equation}
We therefore say that the \emph{infection rate} for a susceptible node with $m$ infected neighbors  is $\lambda \, m$.

%%%%%%%

\subsubsection{Susceptible-Infected-Susceptible (SIS) Model}

Let's consider the somewhat more complicated disease-transmission process in which a node can become susceptible again after becoming infected.  (Alas, permanent recovery is still impossible.)  This process is known as the \emph{susceptible-infected-susceptible (SIS) model}.

We must now introduce stochastic rules for the recovery of infected nodes (i.e., the transition from the infected state to the susceptible state). This transition is usually considered to be a spontaneous process that is independent of the states of neighbors \cite{rom-review2014}. Consequently, each infected node switches to the susceptible state at a constant rate $\mu$. Therefore, in an infinitesimal time interval $dt$, the probability for a node to switch from the infected state $I$ to the susceptible state $S$ is $\mu \,dt$.

%%%%

\subsubsection{Susceptible-Infected-Recovered (SIR) Model}

Another ubiquitous compartmental model is the susceptible-infected-recovered (SIR) model, in which susceptible nodes can still transition to being infected, but infected nodes recover to a state $R$ in which they can no longer be infected \cite{ccc}.  Fatalistic people might let state $R$ stand for ``removed" instead of recovered, but we're going to be more positive than that.

As with the SIS process, two rates define the stochastic dynamics. These are the transmission rate $\lambda$ and the recovery rate $\mu$.  They are both defined as we described above for SIS dynamics, but now the recovery process takes nodes from the $I$ state to the $R$ state rather than returning them to the $S$ state. Interestingly, one can relate the steady-state of the basic SIR model on a network to a bond-percolation process \cite{grassberger83}.  See \cite{kenah2007,trapman2007,lfd2013} for additional discussion of the connection between SIR dynamics and bond percolation.

%%%%%

\subsubsection{More Complicated Compartmental Models}

The contagion models that we have discussed above are interesting to study, and they provide a nice family of tractable examples (including for some analytical calculations, as we will discuss in Sec.~\ref{general}) to examine the effects of nontrivial network structure on dynamics.  They also provide interesting toy situations for biological epidemics, although they are also grossly unrealistic in just about every situation. Nevertheless, they are very illustrative for our goals, and they have significant potential value in demonstrating effects of network structure on dynamical processes that can also occur in more complicated epidemic models.

One step forward is to do similar investigations of more complicated compartmental models on networks.  For example, one can add an ``exposed" compartment to obtain an SEIR model, include age structures, and more \cite{ccc,rom-review2014}.  One can also study \emph{metapopulation} models on networks \cite{colizza2007-meta} to examine both network connectivity and subpopulations with different characteristics, and yet another direction is to examine non-Markovian epidemics on networks to explore the effects of memory \cite{kiss2015b}.

Another option is to throw analytics out the window, be data-driven, and conduct simulations of incredibly detailed and complicated situations while incorporating parameters that are estimated from real data (as well as doing direct data analysis of epidemics).  For example, see Refs.~\cite{colizza2007,balcan2009,tizzoni2013}. Ultimately, it is important to make advancements in all of these approaches, because they complement each other.

%%%%%%%

\subsubsection{Other Uses of Compartmental Models}

A variant set of models involves zombification instead of infections \cite{zombie-book}, and some of the particular details of the models are occasionally slightly different to reflect this different application.  Compartmental models have also been used for various models of social influence \cite{mtbi-reports,karsai-vesp2,Moreno04}, though they are not the most common approach to such topics. (See Sec.~\ref{influence} for a selection of models that were built specifically to study social influence and related phenomena.)

%%%%%%

\subsection{Social Contagions} \label{influence}

Ideas spread along social networks in a manner that appears to be somewhat analogous to biological contagions, and the perceived similarity between social and biological epidemics has led to the use of the term ``contagion" for the spread of social influence \cite{centola2007,yariv2011,jackson2013}. It is common to discuss ideas being ``viral", and some empirical studies have suggested that the spread of ideas in a social network can sometimes be genuinely epidemic-like \cite{yy-epidemic}.  Specifically, an epidemic-like (or ``simple") contagion refers to cases in which---much like with a virus or a disease---exposure to a single source is enough to initiate propagation.  Unlike biological contagions, however, ideas spread in a manner that involves \emph{social reinforcement}: having 100 friends adopt a behavior (or buy a product, join a movement, etc.) can be rather different than if only one friend does so.  Because of social reinforcement, social contagions need not just spread discretely (or even discreetly) from one specific source to another.  This is why they are sometimes called \emph{complex contagions} \cite{centola2007}.

Identifying the causal mechanism of the spread of ideas is more difficult than for the spread of diseases, and development and (especially) validation of models is significantly less mature in social contexts than in biological ones. Even discerning whether or not genuine social influence is occurring in a network is extremely challenging \cite{shalizithomas}, and we find it helpful (for, e.g., the development of models) to illustrate this difficulty based on a data stream that one might encounter in real life.  Suppose that one starts from the empirical observation that the actors represented by the network nodes are adopting or newly exhibiting some sort of behavior at different times (e.g., various actors becoming obese, starting to smoke \cite{fowlerchristakis-fat,fowlerreview,fowlersnark}, or changing their Facebook profile picture into an equal sign \cite{FB-equal,FB-equal2}). It is seemingly common for news outlets to posit such observations as contagions (e.g., for the United Kingdom riots of 2011) \cite{news-riots}, although the same observations can result from one or more of the following effects \cite{shalizithomas}:
\begin{enumerate}
\item{Genuine spread via social influence, though this could also be social learning (as what might be spreading in the network could simply be awareness or knowledge about something rather than strictly some desire to adopt the behavior). Nevertheless, there is something that is genuinely spreading along a network.}
\item{Homophily: Agents tend to adopt the same behavior because they have some common traits that lead to such a propensity.  That is, there is some sort of internal similarity between agents (and that may well even be why some of the network edges exist in the first place), but they happen to be adopting the behavior at different times.}
\item{Environment: There is a common external influence on the agents.  That is, there is some sort of external similarity (or a covariate) that causes agents in a network to adopt the behavior at different times.}
\end{enumerate}
Given observations of agents in a network adopting some behavior at different times, an important goal is to distinguish the relative importances of the above effects.  This is not easy, and control strategies (e.g., legislation) clearly depend on whether the cause of the observations is primarily effect (1), (2), (3), or (most likely) some combination of the three.

To address such issues, it is important to do a lot of data collection (e.g., through surveys, online resources, and other means) along with data analysis and statistics, and the majority of studies of social influence tend to take such a perspective.  However, it is \emph{also} important to develop simple, tractable models of social influence (in which the various effects that resemble putative social influence are, by construction, not confounded with something that genuinely spreads on a network) and to examine such dynamical processes on networks.  Constructing such models yields ``simple" dynamical systems on networks. Efforts to construct simple models of social influence date back at least to the 1970s \cite{degroot,friedkinbook,granovetter78}. They have remained an active topic over the decades, although there has been a rapid acceleration of work on the network aspects of this topic since the publication of Ref.~\cite{watts2002} in 2002.

%%%%%%%%

\subsubsection{Threshold Models}\label{thresh}

Let's start by discussing simple threshold models of social influence, which have a very percolation-like flavor (especially bootstrap percolation \cite{bootstrap}).

In the 1970s, Granovetter posited a simple threshold model of social influence in a fully-mixed population \cite{granovetter78}, and it is natural to consider the effects on network structure on such dynamical processes.  One example is the Watts model \cite{watts2002}, which uses a closely related threshold-based rule for updating states for nodes on a network.  For simplicity, we will restrict our discussion of this and other threshold models to unweighted and undirected networks. However, the Watts model has been generalized to weighted \cite{gleeson-watts-weighted}, directed \cite{gai2010}, temporal \cite{petter-physicaA}, and multilayer \cite{Yagan} networks.

In binary-state threshold models (such as the Watts model, the Centola-Macy model \cite{centola2007}, and others), each node $i$ has a threshold $R_i$ that is drawn from some distribution and which does not change in time. At any given time, each node can be in one of two states: $0$ (inactive, not adopted, not infected, etc.) or $1$ (active, adopted, infected, etc.).  Although a binary decision process on a network is a gross oversimplification of social influence, it can already capture two very important features \cite{oliver1985}: interdependence (an agent's behavior depends on the behavior of other agents) and heterogeneity (differences in behavior are reflected in the distribution of thresholds). Typically, some seed fraction of nodes $\rho(0)$ is assigned to the active state, although that is not always true (e.g., when some nodes have a negative threshold $R_i < 0$).  Depending on the problem under study, one can choose the initially active nodes via some random process (typically, uniformly at random) or with complete malice and forethought.  For the latter, for example, one can imagine planting a rumor with specific nodes in a network.

The states of the nodes change in time according to an update rule.  As with the models in Sec.~\ref{simplecontagions}, one can update nodes either synchronously or asynchronously. The latter, which leads naturally to continuous-time dynamical-systems approximations, will be our main focus. When updating the state of a node in the Watts model, one compares the node's fraction $m_i/k_i$ of infected neighbors (where $m_i$ is the number of infected neighbors and $k_i$ is the degree of node $i$) to the node's threshold $R_i$.  If node $i$ is inactive, it then becomes active (i.e., it is switched to state $1$) if $m_i/k_i \geq R_i$; otherwise, its state remains unchanged.

A similar model to the Watts model is the Centola-Macy model \cite{centola2007}, in which one considers a node's total number $m_i$ of active neighbors rather than the fraction of such neighbors.  (One then writes $m_i \geq R_i$ and makes other similar changes to the formulas that we wrote above.) The special case in which $R_i = R$ for all $i$ in the Centola-Macy model amounts to bootstrap percolation \cite{bootstrap}.

Both the Watts and the Centola-Macy models have a monotonicity property: once a node becomes infected, it remains so forever.  As we will discuss below, this feature is particularly helpful when deriving accurate approximations for the model's global behavior. The Centola-Macy update rule makes it easier than the Watts update rule for hubs to become active, and this can lead to some qualitative differences in dynamics. It is useful to ponder which of these toy models provides a better caricature for different applications.  For posts on Facebook, for example, one can speculate that the number of posts about a topic might make more of a difference than the fraction of one's Facebook friends that have posted about that topic.

Scholars have also studied several more complicated threshold models that have a similar flavor.  For example, one recent interesting threshold model \cite{ruck2013} decomposed the motivation for a node to adopt some behavior as a weighted linear combination of three terms: (1) personal preference, (2) an average of the states of each node's neighbors, and (3) a system average, which is a measure of the current social trend. Another model \cite{prl-bonus} included ``syngergistic" effects from nearby neighbors.  Another study \cite{MelnikChaos13} allowed nodes to be in one of three states: $0$ (inactive), $1$ (active), and $2$ (hyper-active).  In this so-called ``multi-stage'' complex contagion, each node has two thresholds.  An inactive node exerts no influence, an active node exerts some influence, and a hyper-active node exerts both regular influence and some bonus influence.  (A hyper-active node is necessarily active, so state $2$ is a subset of state $1$; however, state $1$ is disjoint from state $0$.) In such a multi-stage generalization of the Watts model, a node updates its state when its ``peer pressure" $P = [l_1 + \beta l_2]/k$ equals or exceeds a threshold. (The multi-stage version of the Centola-Macy model has a peer pressure of $P = [l_1 + \beta l_2]$.) The number of neighbors in state $i$ is $l_i$, and $\beta$ is the bonus influence.  In \cite{MelnikChaos13}, a node whose peer pressure equals or exceeds the first threshold $R_j^{(1)}$ achieves state $1$, and a node whose peer pressure equals or exceeds the second threshold $R_j^{(2)} \geq R_j^{(1)}$ achieves state $2$.  Note that this multi-stage complex contagion model is still monotonic.  Some studies have considered non-monotonic generalizations of the Watts and similar threshold models---e.g., by including ``hipster" nodes that become inactive if too many of their neighbors are active \cite{DoddsDeckerPRL}. It is also interesting to study interactions between biological and social contagions \cite{funk-review2010}.

%%%%%%%

\subsubsection{Other Models}

Although the threshold models of social influence that we described above have the advantage of being mathematically tractable (at least when suitable approximations hold, as we discuss Sec.~\ref{general}) and providing a nice (and convenient) caricature of adoption behavior, they are way too simplistic. For example, the aforementioned threshold models do not consider specific signals from nodes (e.g., an individual tweeting about the same thing multiple times), which is a ``more microscopic" aspect of human behavior.  Social reinforcement can arise from multiple friends adopting the same or similar behavior, but it can also arise from the same person sending multiple signals.  Moreover, the threshold models are ``passive" in the sense that the only signal that ever arises from a node is whether or not it exhibits a behavior (i.e., what state it is in) and all of the dynamics focus entirely on whether a node gets enough influence (purely from what state other nodes are in) to adopt a new behavior. Reference~\cite{borge-intfire} tried to address this situation by adopting an idea from neuroscience \cite{termanbook} by supposing that each person is an integrate-and-fire oscillator, where the ``firing" corresponds to actively sending a signal (e.g., sending a tweet).  This can then lead to other nodes adopting the behavior and sending their own signals. Recent models of tweeting have examined the spreading of ideas and memes, with an emphasis on the competition between memes for the limited resource of user attention \cite{Weng12,Gleeson13b,Gleeson15}.

Approaches to modeling social influence and social learning besides threshold models also date back at least to the 1970s, and some of the models that have been studied are also analytically tractable (although they have a rather different flavor from threshold models).  For example, in the DeGroot model \cite{degroot}, individual $j$ has opinion $y_j$, and the discrete-time opinion dynamics satisfy the equation
\begin{equation}
	{\bf y}(t + 1) = {\bf W}{\bf y}(t)\,, \qquad t = 0\,, 1\,,2\,, \ldots \,,
\end{equation}
where {\bf W} is a row-stochastic weight matrix (so that $\sum_j w_{ij} = 1$ for all $i$), and the matrix element $w_{ij}$ (including the case $i = j$) represents the influence of node $j$ on node $i$.  Friedkin and coauthors (among others) have generalized the DeGroot model in many ways \cite{friedkinbook}.

There are, of course, many other models (see, e.g., the discussion in the introduction of Ref.~\cite{MelnikChaos13}), so our treatment should not be viewed as even remotely exhaustive.  See Ref.~\cite{yamir-jcn2013,rom-review2014,yariv2011} for additional discussions.

%%%%%%%

\subsection{Voter Models} \label{votes}

Another well-known dynamical system that is often studied on networks is the so-called \emph{voter model} \cite{loreto2009}.  Voter dynamics were first considered by Clifford and Sudbury \cite{cliff1973} in the 1970s as a model for species competition, and the dynamical system that they introduced was dubbed the ``voter model" by Holley and Liggett a couple of years later \cite{holley1975}. Voter dynamics are fun and versatile (and are very interesting to study on networks), though it is important to ask whether one can ever genuinely construe the voter model (or its variants) as a model for voters \cite{ramasco2014}. 

The standard (or ``direct'') voter model is defined on a network as follows. Each node has a binary variable that can either be in state $+1$ or in state $-1$.  (For example, the former might represent the US Democratic party, and the latter might represent the US Republican party.)  At every discrete time step, one selects some node $i$ uniformly at random, and node $i$ then adopts the opinion $s_j$ of one of its neighbors $j$ (which is selected uniformly at random from among all of $i$'s neighbors).  If $i$ and $j$ were already voting in the same way before the time step, then no change occurs.  We remark that one can map the basic voter model to a model of random walkers that coalesce when they encounter each other \cite{cliff1973,holley1975}. An alternative to the direct voter model is the ``edge-update'' voter model \cite{Suchecki05}, in which one chooses an edge (rather than a node) uniformly at random at each time step. If the opinions of the nodes at the two ends of the chosen edge are different, then we randomly select one of the nodes, and that node adopts the opinion of the other node. As discussed in Ref.~\cite{loreto2009}, there are a wealth of studies and results (including mathematically rigorous ones) on the voter model.

The original voter model is of course a gross oversimplification of reality, but it is analytically tractable and provides a foundation for numerous interesting generalizations. Indeed, there are a large number of variants of the original voter model, and many of them provide fodder for wonderfully snarky jokes (e.g., see below), and this is especially true if one chooses to label the opinions of nodes with terminology such as ``infected'' (as is sometimes tempting in political discussions). These models also grossly oversimplify reality, but they are fascinating, they are sometimes mathematically tractable (depending on the network structure under consideration), and they can even yield insights that are legitimately interesting for applications. For example, whether consensus is reached and how long it takes to reach consensus (or other equilibrium states) depends on both the specific dynamics and on the network on which those dynamics occur.  For the direct voter model on configuration-model networks with a power-law degree distribution, the mean consensus time scales linearly with the number $N$ of nodes in the network if the exponent $\gamma$ of the degree distribution exceeds $3$, whereas it scales sublinearly with $N$ if $\gamma\le 3$ \cite{sood05,sood08}. By contrast, the edge-update voter dynamics can have different asymptotic properties: for example, in BA networks (which have power-law degree distributions as $N \rightarrow \infty$), the consensus time depends linearly on $N$ for any exponent \cite{Castellano05}.

One nice variant voter model is a ``constrained" voter model \cite{vazquez2003} (see also the more general ``political positions process'' in \cite{itoh1998} and recent work such as \cite{lanchier2012}) in which nodes can be in one of three states (Left, Right, and Center).  All interactions in the constrained voter model involve centrists, as extremists refuse to talk to each other.  By considering this model on a complete graph and thereby examining the mean-field limit (see Sec.~\ref{MFPA} for a discussion of mean-field and related approximations) in which the voters are perfectly mixed, V\'azquez et al. \cite{vazquez2004} derived probabilities, which depend on the initial conditions, of reaching a consensus in one of the three states or of achieving a mixture of the two extremist states.  Another interesting variant of the voter model is the ``vacillating" (two-state) voter model \cite{lambred2007}, in which the node $i$ that has been selected examines the states of two of its neighbors, and it changes its state if either of them is different from its own state \footnote{As an example of a snarky joke, one might imagine that this model is more realistic in some countries than in others. Identification of any such countries is left as an exercise for the diligent reader.}. See Ref.~\cite{loreto2009} for discussions of many more generalizations of the voter model.

%%%%%%

\subsection{Interlude: Asynchronous Versus Synchronous Updating} \label{async}

Before proceeding to our discussion of other types of dynamical processes on networks, it is useful to pause and examine how to implement the update rules in discrete-state dynamics.

When simulating (stochastic or deterministic) discrete-state dynamics on networks, it is necessary to decide on a
a method for choosing which nodes to update and when to update them. Some dynamical processes are defined in a way that is simple to simulate numerically. The voter model described in Sec.~\ref{votes}, for example, is explicitly defined in terms of discrete time steps, and one node is chosen uniformly at random to update in each time step. This is a form of \emph{asynchronous updating}, which goes by that name because individual nodes are updated independently, such that the new state of a node becomes visible to its neighbors before they attempt to update their own states. One can also employ asynchronous updating for the Watts threshold model described in Sec.~\ref{thresh}.  One again chooses a node uniformly at random in each time step and---if it is in the inactive state---one compares the fraction of its active neighbors to its threshold to determine if it becomes active. Alternatively, one can choose to update the states of all nodes simultaneously in each time step; this is called \emph{synchronous} (or \emph{parallel}) updating. When updating in this way, nodes change their states based on the states of their neighbors from the previous time step. If one is updating using discrete time steps (as is common in computer simulations), then one can construe the above synchronous and asynchronous schemes as limiting cases of a more general update scheme in which a fraction $f$ of the nodes are chosen uniformly at random in each time step and one then updates these particular nodes synchronously. If $f=1/N$, then (on average) one node is updated in each time step, giving the asynchronous methods that we described above (as used, for example, in voter models). The choice $f=1$ gives synchronous updating.

Synchronous updating has the advantage of allowing fast simulations, but asynchronous updating leads to more gradual changes, because only one node is update per time step, so the fraction of active nodes changes by at most $1/N$ in a time step. Consequently, asynchronous updating leads---in the limit in which a vanishingly small fraction of the nodes (e.g., $f = 1/N$) are updated in each discrete time step---to dynamics that are accurately described by a set of coupled differential equations. For certain classes of dynamics (e.g., the monotonic dynamics of Sec.~\ref{thresh}, the steady-state (i.e., $t\to\infty$) limits obtained using either synchronous or asynchronous updating schemes are identical, but this need not be true in general.  Moreover, the finite-time dynamics are clearly different for asynchronous versus synchronous updating even when the $t\to\infty$ limits are identical.

For stochastic dynamical processes (such as the biological contagion models of Sec.~\ref{simplecontagions}), which are defined in terms of hazard rates, some care is needed in the implementation of an update rule in computational simulations \cite{holme-tute2014}. If a given node $i$ has a rate $F_i$ (i.e., a so-called ``hazard rate'') for changing states, then it has a probability of $F_i \, dt$ of changing its state during an infinitesimal time interval $dt$. The ``infinitesimal'' part of this definition is important: it requires that the discrete time step $dt$ of simulations is very small. In practice, $dt$ should be sufficiently small so that only one (or at most a few) nodes are updated in each step. Implementing an update rule in this way ensures that the underlying processes are faithfully reproduced. For example, the length of time $T$ that a node spends in its current state before being updated (assuming that no neighbors are updated during this time) should be exponentially distributed. To see that this is reproduced in simulations, note that for each time step of length $dt$, the probability of node $i$ not changing states is $1-F_i \, dt$. Because there are $T/dt$ discrete time steps in the interval $[0,T]$, the probability that the node survives until time $T$ without changing state is the product of the survival probabilities in each step:
\begin{equation}
	\text{Prob(survival until at least $T$)} = \left( 1- F_i \, dt\right)^\frac{T}{dt}\,.
\end{equation}
In the $dt\to 0$ limit, this yields the exponential distribution of survival times (as expected for a Markov process \footnote{Naturally, it is also relevant to consider the effects of memory on dynamical processes on networks \cite{rosvall2014,kiss2015b}.}, where the probability of changing state depends only on the current state of the system):
\begin{equation}
	\lim_{dt\to 0} \left( 1- F_i \, dt\right)^\frac{T}{dt} = \exp\left(-F_i T\right)\,.
\end{equation}

One can use the interpretation of stochastic transition rates in terms of survival times to consider alternative asynchronous updating methods, such as Gillespie or Kinetic Monte Carlo algorithms \cite{Gillespie77,boguna2013}, which are event-driven rather than using equally-spaced time steps. Although they are less straightforward to code, Gillespie algorithms can considerably accelerate simulation times for certain dynamical processes, and we expect their use for stochastic dynamics (including for non-Markovian dynamics \cite{boguna2013} and dynamical processes on temporal networks \cite{vestergaard2015}) to become increasingly popular.

%%%%%%%%

\subsection{Coupled Oscillators} \label{kuramoto}

Coupled oscillators are a very heavily studied type of dynamical system, and associating each oscillator to a node of a network allows one to investigate how nontrivial connectivity affects collective phenomena such as synchronization \cite{arenas-review}.  Perhaps the most famous model of coupled oscillators is the \emph{Kuramoto model} \cite{kuramoto1984,strogatz2000,arenas-review,bonilla-review,kuramoto2014-review} of phase oscillators. It is one of the canonical models to use in the study of \emph{synchronization} \cite{scholarsync,arkadybook}, which refers to an adjustment of rhythms of oscillating objects due to their (possibly weak) interactions with each other.  The Kuramoto model is also one of the most popular dynamical systems to study on networks. Because each node is an oscillator, it provides an interesting contrast to the contagion and voter models that we discussed above.

In the Kuramoto model, each node $i$ has an associated phase $\theta_i \in [0,2\pi)$ whose dynamics are governed by
\begin{align} \label{kuramoto-equ}
	\dot{\theta}_i = \omega_i + \sum_{j = 1}^N b_{ij} A_{ij} f_{ij}(\theta_j - \theta_i)\,, \qquad i \in \{1,\ldots, N\}\,,
\end{align}
where the the natural frequency $\omega_i$ of node $i$ is typically drawn from some distribution $g(\omega)$ (though it can also be deterministic), ${\bf A} = [A_{ij}]$ is the adjacency matrix of an unweighted network, $b_{ij}$ gives the coupling strength between oscillators $i$ and $j$ (so that $b_{ij}A_{ij}$ gives an element of a weighted network), and $f_{ij}(y)$ is some coupling function that depends only on the phase difference between oscillators $i$ and $j$.

Equation (\ref{kuramoto-equ}) is much more general than the traditional Kuramoto model, for which $f_{ij}(y)$ is the same function $f(y)$ for all node pairs, the coupling function is $f(y) = \sin(y)$, and $b_{ij} = b$ for all node pairs.  The traditional networks on which to study the Kuramoto model have either all-to-all coupling or nearest-neighbor coupling, but it is both popular and very interesting to examine the Kuramoto model on networks with more general architectures. The properties of $g(\omega)$ have a significant effect on the dynamics of equation (\ref{kuramoto-equ}). For example, it is important whether or not $g(\omega)$ has compact support, whether or not it is symmetric, and whether or not it is unimodal.  In traditional studies of the Kuramoto model, $g(\omega)$ is unimodal and symmetric about some mean frequency $\Omega$.  The original Kuramoto model also uses all-to-all coupling.

To study the original Kuramoto model, one can consider deviations from the mean frequency by transforming to a rotating frame via $\nu_i := \omega_i - \Omega$.  This allows one to directly see which oscillators are faster are than the mean and which ones are slower than the mean. One then defines the complex ``order parameter''
\footnote{In statistical physics, an \emph{order parameter} is a quantity (e.g., a scalar) that summarizes a system and is used to help identify and measure some kind of order \cite{sethnabook}.} \cite{strogatz2000,arenas-review}
\begin{align} \label{order}
	r(t)e^{i\psi(t)} := \frac{1}{N}\sum_{j = 1}^N e^{i \theta_j (t)}\,,
\end{align}
where $r(t) \in [0,1]$ measures the coherence of the set of oscillators and $\psi(t)$ gives their mean phase.   The parameter $r$ quantifies the extent to which the oscillators exhibit a type of synchrony known as \emph{phase-locking}, which is a form of synchrony in which (as the name implies) the phase differences between each pair of oscillators have the same constant value. When $r =1$, the oscillators are {phase-locked}, and they are completely incoherent when $r = 0$. However, these extreme situations only occur in the thermodynamic ($N \rightarrow \infty$) limit. In practice, $r \approx 1$ (rather than $r = 1$) for a finite number of synchronized oscillators, and $r \approx 0$ for a finite number of completely incoherent oscillators. The finite-size fluctuations have a size of $O(1/\sqrt{N})$ \cite{strogatz2000}.  Additionally, note that one can affect synchronization properties in interesting ways by perturbing the Kuramoto model with noise (which can either promote or inhibit synchrony, depending on the precise details) \cite{yiming-mason}.

When placing Kuramoto oscillators on a network, one can then ask the usual question: How does nontrivial network topology (i.e., connectivity) affect the synchronization dynamics of the oscillators \cite{arenas-review,kuramoto2014-review}?  In addition to numerical simulations, one can conduct analytical investigations using generalizations of the order parameter in (\ref{order}) along with concomitant calculations (see, e.g., \cite{ichi2004,ichi2005,lee2005,restrepo2005}) that are more intricate versions of what has been used in studies of the original Kuramoto model \cite{kuramoto1984,strogatz2000,arenas-review}.

A particularly interesting investigation of Kuramoto dynamics on networks was a recent examination of \emph{explosive synchronization} \cite{arenas-kaboom} that was motivated by studies of explosive percolation (see the discussion in Sec.~\ref{kaboom}) \cite{raissa-kaboom}.  Reference~\cite{arenas-kaboom} elucidated a situation that can lead to a genuinely ``explosive" (i.e., discontinuous or ``first-order") phase transition in a set of interacting Kuramoto oscillators,
\begin{align} \label{kuram2}
	\dot{\theta}_i = \omega_i + b\sum_{j = 1}^N A_{ij} \sin(\theta_j - \theta_i)\,, \qquad  i \in \{1,\ldots, N\}\,,
\end{align}
on a family of networks (see Ref.~\cite{moreno2006erba} for a precise specification) that interpolates between Barab\'asi-Albert (BA) networks in one limit and Erd\H{o}s-R\'enyi (ER) networks in the other limit.  This family of networks is parametrized by one parameter, which we denote by $\alpha$.  One obtains a BA network when $\alpha = 0$ and an ER network when $\alpha = 1$.  Suppose that the oscillator frequency $\omega_i \propto k_i^\beta$ (where $\beta > 0$) is positively correlated with node degree.  In contrast to most studies of Kuramoto oscillators, these natural frequencies are deterministic rather than chosen randomly from a (nontrivial) distribution. Plotting $r(t)$ from equation (\ref{order}) versus the coupling strength $b$ illustrates a phase transition that appears to become discontinuous in the $\alpha \rightarrow 0$ limit.  The positive correlation between node degree and the natural frequency of the oscillators seems to lead to a positive feedback mechanism that makes it possible to have a discontinuous phase transition.

To verify that one can truly obtain a discontinuous phase transition (and hence a genuinely ``explosive" synchronization transition \footnote{Recall from Sec.~\ref{kaboom} that so-called ``explosive" percolation is actually a steep but continuous transition \cite{oliver-science,oliver-aap,dorog2010}.}), let's consider a star graph.  In a star network, there is a central hub node that is adjacent to all other nodes, which are each only adjacent to the hub.  Suppose that there are $N = K+1$ nodes, so that the hub has degree $K$ and the $K$ leaf nodes each have degree $1$.  Denote the natural frequency of the hub oscillator by $\omega_h$, and let each leaf node have a natural frequency of $\omega$.

We let $\varphi(t) = \varphi(0) + \Omega t$, where
\begin{equation}
	\Omega = \frac{K\omega + \omega_h}{K+1}
\end{equation}
is (as usual) the mean frequency of the oscillators. We take $\varphi(0) = 0$ without loss of generality, because we can uniformly shift the phases of all oscillators.  We thus transform the angular variables as follows:
\begin{align}
	\varphi_h &:= \theta_h - \Omega t\,, \qquad \mathrm{(hub)}\,,\notag \\
	\varphi_j &:= \theta_j - \Omega t\,, \qquad j\in \{1,\ldots,K\}\,,
\end{align}
where we have labeled the $(K+1)$th node using ``$h$'' because it is the hub. The equations of motion (\ref{kuram2}) thus become
\begin{align}\label{hubhub}
	\dot{\varphi}_h &= (\omega_h - \Omega) + b\sum_{j = 1}^K \sin(\varphi_j - \varphi_h)\,, \notag \\
	\dot{\varphi}_j &= (\omega - \Omega) + b\sin(\varphi_h - \varphi_j)\,, \qquad j\in \{1,\ldots,K\}\,.
\end{align}

As usual, we define the order parameter using equation (\ref{order}) and hence write
\begin{align} \label{order2}
	r(t)e^{i\psi(t)} := \frac{1}{K+1}\sum_{j = 1}^{K+1} e^{i \varphi_j (t)} \equiv \langle e^{i\varphi}\rangle\,,
\end{align}
where we note that one can express the complex order parameter in terms of an ensemble average over the oscillators. In our current coordinates (a rotating reference frame), the mean oscillator phase is $0$, so $\psi = \varphi(0) = 0$, and separately equating real and imaginary parts in equation (\ref{order2}) yields
\begin{align}
	{r(t)} = \frac{1}{K+1}\sum_{j = 1}^{K+1}\cos\varphi_j(t) \equiv \langle \cos \varphi(t) \rangle\,, \qquad
	0 = \frac{1}{K+1}\sum_{j = 1}^{K+1}\sin\varphi_j(t) \equiv \langle \sin \varphi(t) \rangle\,.
\end{align}

%{\bf map: I'm having a bit of notation inconsistency on the angle names, so I'll need to come back to this (because of this example versus the above); James, please double-check me for the notation in the Kuramoto section; we will want to check this again at the page proof stage}

We multiply equation (\ref{order2}) by $e^{-i\varphi_h(t)}$ to obtain
\begin{align}\label{equate0}
	re^{i(\psi - \varphi_h)} = re^{-i\varphi_h} = \frac{1}{K+1}\sum_{j = 1}^{K+1} e^{i (\varphi_j - \varphi_h)}\,,
\end{align}
which implies that	
\begin{align}\label{equate}	
	r\cos \varphi_h - i r\sin \varphi_h &= \frac{1}{K+1}\sum_{j = 1}^{K+1} \cos(\varphi_j - \varphi_h) + i \frac{1}{K+1}\sum_{j = 1}^{K+1}\sin(\varphi_j - \varphi_h) \notag \\
	&=  \frac{1}{K+1}\left[1 + \sum_{j = 1}^{K} \cos(\varphi_j - \varphi_h)\right] + i \frac{1}{K+1}\sum_{j = 1}^{K}\sin(\varphi_j - \varphi_h)\,,
\end{align}
where we have separated the hub term (with $j = K + 1 = h$) from the other terms in the sums. (Note that we are now suppressing the explicit indication of time-dependence for quantities such as $r(t)$ and $\varphi(t)$.) We separately equate the real and imaginary parts of equation (\ref{equate}), and the latter yields
\begin{align}
	-r\sin\varphi_h = \frac{1}{K+1}\sum_{j = 1}^K \sin(\varphi_j - \varphi_h)\,,
\end{align}
which we insert into the first equation in (\ref{hubhub}) to obtain \footnote{Note that the analog of equation (\ref{sign}) in Ref.~\cite{arenas-kaboom} has a sign error.}
\begin{align}\label{sign}
	\dot{\varphi}_h = (\omega_h - \Omega) - b(K+1){r}\sin(\varphi_h)\,,
\end{align}
where the second term is a mean-field coupling term because all interactions with other oscillators depend only on the ensemble average.

The hub oscillator is phase-locked when $\dot{\varphi}_h = 0$ (i.e., when the relative phases are constant), which occurs when
\begin{equation}
	\sin(\varphi_h) = \frac{\omega_h - \Omega}{b(K+1){r}}\,.
\end{equation}
For leaf oscillators to be phase-locked, we need $\dot{\varphi}_j = 0$ for all $j \in \{1,\ldots, K\}$.  We thus require that
\begin{align}
	\cos \varphi_j = \frac{1}{b}\left \{	 (\Omega - \omega)\sin\varphi_h \pm \sqrt{\left[1 - \sin^2(\varphi_h)\right]\left[b^2 - \left(\Omega - \omega\right)^2\right]}\right \}\,, \qquad j \in \{1,\ldots, K\}\,,
\end{align}
which is valid as long as $\Omega - \omega \leq b$.  We lose phase-locking at a critical coupling $b = b_c := \Omega - \omega$.  For example, if $\omega_h = K$ and $\omega = 1$ (i.e., when each oscillator frequency is equal to the degree of the associated node), we obtain
\begin{align}
	\Omega = \frac{2K}{K+1}\,, \qquad b_c = \frac{K-1}{K+1}\,.
\end{align}	
The critical value ${r}_c$ of the order parameter ${r}$ is then
\begin{align}
	{r}_c = r(b = b_c) = \left.\frac{\cos\varphi_h + K\cos \varphi_j}{K+1}\right|_{b = b_c} = \frac{K}{K+1} >0\,.
\end{align}
Because ${r}_c > 0$, we obtain a vertical gap (i.e., a discontinuous synchronization transition) in the phase diagram (i.e., bifurcation diagram) of ${r}$ versus $b$. When $K \rightarrow \infty$, the critical value $r_c \rightarrow 1$, so there is a discontinuous (i.e., ``explosive") phase transition from no synchrony (i.e.,. complete incoherence) to complete synchrony in the thermodynamic limit. Kaboom!

%%%%%%

\subsection{Other Dynamical Processes}

Numerous other dynamical systems have also been studied on networks, and obviously many others can be.  In this section, we briefly mention a few of them.  As with most of our other discussions, we are not being even remotely exhaustive.

General ideas for dynamical systems on networks (see Sec.~\ref{master} for an example of one type of methodology) have been used to examine stability for a wealth of both continuous and discrete dynamical systems on networks \cite{mackay1995,pecoracarroll1998,gade2000,pereira2013}. Such ideas have been applied to investigate numerous phenomena, including synchronization of chaotic systems (such as R\"ossler circuits \cite{fink2000}) on networks. In addition to stability, there are also many studies  on the control of dynamical systems on networks \cite{mario-review2010,motter-expo,liu-control2011,cowan2012,selley2014,ruths2014}.

As we have discussed at length, there is a lot of work that examines ODEs on networks, but comparatively little work has addressed PDEs on networks. Thankfully, the amount of scholarship on both linear and nonlinear PDEs on networks is starting to increase (see, e.g., \cite{smilansky2006,herrmann2014,ide2014,malbor2014}), and there are many exciting avenues to pursue.  For example, it is intuitively sensible to study models of vehicular (and other) traffic flow on networks \cite{trafficbook}, and studying shock-forming PDEs (like Burger's equation), which have often been employed for models of traffic flow \cite{whitham}, on networks should be really interesting.

A family of examples that is related to some of our previous discussions (see Sec.~\ref{thresh}) are models of financial contagions and systemic risk in banking networks in terms of threshold models of complex contagions but with additional complications \cite{gai2010,may2011,hurdgleeson,doynebipartite}. One can reduce the simplest models to threshold models on directed networks, but the incorporation of weighted edges (e.g., to represent the values of interbank loans) is important for the development of increasingly realistic models.  It is also essential to incorporate other features of financial contagions.

Coordination games on networks---a subtopic of the much larger topic of games on networks (see \cite{jackson2010,jackson2014} for extensive discussions of an enormous variety of games)---are also directly related to threshold models \cite{yariv2011}. Consider the example of technology adoption, and suppose that there are two choices (i.e., game strategies) $A$ and $B$ that represent, for example, technologies for communicating with friends (e.g., mobile-phone text messaging versus Facebook messaging). The game is played by the nodes on a network, where each node adopts one of the possible strategies. If two neighbors on the graph each choose strategy $A$, then they each receive a payoff of $q$; if they both choose strategy $B$, they each receive $1-q$; and if they choose opposite strategies, they receive $0$ payoff because they cannot communicate with each other. If $q>1/2$, then strategy $A$ is construed as representing the superior technology. Consider a network in which all nodes initially play strategy $B$, and a small number of nodes begin adopting strategy $A$. If we apply best-response updates to the nodes, they end up adopting strategy $A$ only when enough of their network neighbors have already adopted $A$ (as in the Watts threshold model). Specifically, suppose that $m$ of the $k$ neighbors of a given node are playing strategy $A$ and the remaining $k-m$ neighbors of the node are playing strategy $B$. If a node plays $A$, then his/her payoff is $m q$; however, if he/she plays $B$, then his/her payoff is $(k-m)(1-q)$. Comparing the payoffs in the two situations, we see that the node should play strategy $B$ until the fraction $m/k$ of his/her neighbors who are playing $A$ is at least $1-q$  \cite{EasleyKleinberg}. The spreading of technology $A$ thus proceeds as a complex contagion precisely as specified in the Watts model. There exist rigorous mathematical results on the speed of adoption cascades in such coordination games on various network topologies \cite{Montanari10,EasleyKleinberg}. See Sec.~\ref{subsubcascade} for a discussion of cascades in complex contagions.

Given the natural network description of neuronal systems, it is crucial to investigate simple models of neural signal propagation on networks \cite{termanbook} in order to illuminate the effects of network structure on dynamics (and, ideally, on functional behavior).  Understandably, it is desirable to understand the synchronization properties of such models in networks with different topologies, and some investigations of Kuramoto models on networks have had neuronal applications in mind. It is necessary, of course, to consider dynamical processes that are more directly relevant for neuroscience applications. Neuronal signal models that have been studied on networks include integrate-and-fire models \cite{brunel2002}, Hindmarsh-Rose oscillators \cite{dhamala2004}, Fitzhugh Nagumo models \cite{cakan2013}, and more. In neuroscience applications, it is also important to investigate the effects of delay (e.g., due to  propagation along longer neurons) \cite{dhamala2004,cakan2013}.

Oscillator synchrony on networks also plays a role in animal behavior, such as in cattle synchrony (e.g., it is supposedly helpful for cattle welfare when cows lie down at the same time) \cite{cow-sync}.  Moreover, the model for cow behavior in Ref.~\cite{cow-sync} exhibits an interesting feature that has also recently been noted more generally for dynamical systems on networks \cite{motter-weirdlesssink}: it is possible that increasing the coupling strength of edges or increasing the number of coupling edges can \emph{lower}, rather than raise, the amount of synchrony.  The latter case (i.e., reducing synchrony by adding new edges) is a direct analog of the Braess paradox from traffic systems \cite{braess,braess-shl}.

Many other types of dynamical processes have also been studied on networks.  For example, there is an overwhelming abundance of research on dynamical processes on networks in ecology \cite{pascual1995,bascompte2013,rough2014}. The study of Boolean networks is also very popular, and such networks have been used for models in subjects such as genetic circuits \cite{kauffman1969,chaves2005,pomerance2009}, opinion dynamics \cite{loreto2009}, and more.  Chemical-reaction networks have also been studied for a long time and from many perspectives (see, e.g., \cite{feinberg1979,craciun2005,craciun2006}).

%%%%%%

\section{General Considerations} \label{general}

Now that we have discussed several families of specific models as motivation (and because they are interesting in their own right), we present some general considerations.  We alluded to several of these ideas in our prior discussions.

%%%%

\subsection{Master Stability Condition and Master Stability Function} \label{master}

In this section, we derive a so-called \emph{master stability condition} and a \emph{master stability function}, which make it possible to relate the qualitative behavior of a dynamical system on a network to network structure via eigenvalues of the associated adjacency matrix \cite{pecoracarroll1998}.  (One can also, of course, express such results in terms of the eigenvalues of graph-Laplacian matrices.) Computation of matrix spectra is easy, so this provides a convenient means to obtain necessary and sufficient conditions for the linear stability of equilibria, periodic orbits, or other types of behavior. The use of such techniques is therefore very common in the investigation of phenomena such as synchronization in networks of coupled oscillators \cite{arenas-review}. In our discussion, we closely follow (parts of) the presentation in Newman \cite{newman2010}, which illustrates these ideas in the context of continuous dynamical systems. As usual, one can consider more general situations than what we will present \cite{arenas-review}.

Let's suppose that each node $i$ is associated with a single variable $x_i$.  We use ${\bf x}$ to denote the vector of variables. Consider the continuous dynamical system
\begin{align}\label{dyn1}
	\frac{d x_i}{d t}\equiv	\dot{x}_i = f_i(x_i) + \sum_{j = 1}^N A_{ij} g_{ij}(x_i,x_j)\,, \qquad i \in \{1, \ldots, N\}\, ,
\end{align}
where ${\bf A} = [A_{i j}]$ is the network adjacency matrix and $g_{i j} (x_i,x_j)$ gives the effect of network neighbors on each others' dynamics. As usual, the equilibrium points for equation (\ref{dyn1}) \cite{stro1994,guckholmes} satisfy $x_i = 0$ for all nodes $i$.  To determine the local stability of these points, we (of course) do linear stability analysis: let $x_i = x_i^* + \epsilon_i$ (where $|\epsilon_i| \ll 1$) and take a Taylor expansion.  We will assume that the network represented by the adjacency matrix ${\bf A}$ is static, so one can clearly be much more general than what we do for our presentation.  (There are also other ways in which one can be more general.) For each $i$, we obtain
\begin{align}
	\dot{x}_i &= \dot{\epsilon}_i  = f_i(x_i^* + \epsilon_i) + \sum_{j = 1}^N A_{ij} g_{ij}(x_i^* + \epsilon_i,x_j^* + \epsilon_j)\,, \notag \\
		&= f_i(x_i^*) + \sum_{j = 1}^N A_{ij} g_{ij}(x_i^*,x_j^*) + \left.\epsilon_i {f}_i'\right|_{x_i = x_i^*} + \epsilon_i\sum_{j = 1}^N A_{ij} \left.\frac{\partial g_{ij}}{\partial x_i}\right|_{x_i = x_i^*,x_j = x_j^*} + \sum_{j = 1}^N \epsilon_j A_{ij}\left.\frac{\partial g_{ij}}{\partial x_j}\right|_{x_i = x_i^*,x_j = x_j^*} + \dots \notag \\
			&\equiv \alpha_1 + \alpha_2 + \alpha_3 + \alpha_4 + \alpha_5 + \dots\,,
\end{align}
where $f_i' := \frac{df_i}{dx_i}$ and the $\alpha_l$ terms are defined in order from the corresponding terms in the preceeding equation.  Because ${\bf x}^*$ is an equilibrium, it follows that $\alpha_1 + \alpha_2 = 0$.  The terms $\alpha_3$ and $\alpha_4$ are linear in $\epsilon_i$, and $\alpha_5$ is linear in $\epsilon_j$.  We are doing linear stability analysis, s we neglect all higher-order terms.

To simplify notation, we define
\begin{align}
	a_i &:= \left.{f}_i'(x_i)\right|_{x_i = x_i^*}\,, \notag \\
	b_{ij} &:=  \left.\frac{\partial g_{ij}(x_i,x_j)}{\partial x_i}\right|_{x_i = x_i^*,x_j = x_j^*} \,, \notag \\
	c_{ij} &:= \left.\frac{\partial g_{ij}(x_i,x_j)}{\partial x_j}\right|_{x_i = x_i^*,x_j = x_j^*}\,. \label{was25}
\end{align}
That is,
\begin{equation}
	\dot{{\bf \epsilon}} = {\bf M \epsilon} + \dots \,,
\end{equation}
where ${\bf M} = [M_{ij}]$ and	
\begin{equation}
	M_{ij} = \delta_{ij}\left[a_i + \sum_kb_{ik}A_{ik}\right] + c_{ij}A_{ij}\,.
\end{equation}
Assuming that the matrix $\mathbf{M}$ has $N$ unique eigenvectors (which need not always be the case \cite{guckholmes}, although it will usually be if we are away from a bifurcation point) and is thus diagonalizable, we can expand
\begin{equation}
	{\bf \epsilon} = \sum_{r = 1}^N \alpha_r(t) {\bf v}_r\,,
\end{equation}	
where ${\bf v}_r$ (with corresponding eigenvalue $\mu_r$) is the $r$th (right) eigenvector of the matrix ${\bf M}$.  It follows that
\begin{align} \label{diag}
	\dot{\bf \epsilon} = \sum_{r = 1}^N\dot{\alpha}_r{\bf v}_r = {\bf M \epsilon} = {\bf M}\sum_{r = 1}^N\alpha_r(t){\bf v}_r = \sum_{r = 1}^N\alpha_r(t){\bf Mv}_r = \sum_{r = 1}^N\mu_r\alpha_r(t){\bf v}_r\,.
\end{align}
Separately equating the linearly independent terms in equation (\ref{diag}) then yields $\dot{\alpha}_r = \mu_r\alpha_r$, which in turn implies that $\alpha_r(t) = \alpha_r(0)\exp(\mu_r t)$.  As usual for dynamical systems \cite{stro1994,guckholmes}, we obtain local asymptotic stability if $\mathrm{Re}(\mu_r) < 0$ for all $r$, instability if any $\mathrm{Re}(\mu_r) > 0$, and a marginal stability (for which one needs to examine nonlinear terms) if any $\mathrm{Re}(\mu_r) = 0$ for some $r$ and none of the eigenvalues have a positive real part.

As an example, let's consider a (significantly) simplified situation in which every node has the same equilibrium location: i.e., $x_i^* = x^*$ for all nodes $i$.  (This arises, for example, in the SI model of a biological contagion.) We will also assume that $f_i \equiv f$ for all nodes and $g_{ij} \equiv g$ for all node pairs.  This is also a major simplification, but it is employed in the overwhelming majority of studies in the literature, predominantly because it is convenient and because it is still hard to do analytical studies of dynamical systems on networks even with these simplifications.  After applying the simplifications, we can write
\begin{align}\label{simp}
	f(x^*) + \sum_{j = 1}^NA_{ij}g(x^*,x^*) = f(x^*) + k_ig(x^*,x^*) = 0\,,
\end{align}
where we recall that $k_i$ is the degree of node $i$.  (Recall as well that we are considering unweighted and undirected networks.) Equation (\ref{simp}) implies that either all nodes have the same degree (i.e., that our graph ``$z$-regular," where $z$ is the degree) or that  $g(x^*,x^*) = 0$.  We do not wish to restrict the network structure this severely, so we will suppose that the latter condition holds.  It follows that $f(x^*)=0$, so the equilibria of the coupled equation (\ref{dyn1}) in this simplified situation are necessarily the same as the equilibria of the intrinsic dynamics that are satisfied by individual (i.e., uncoupled) nodes.  This yields a simplified version of the notation from equation (\ref{was25}):
\begin{align}
	a_i &\equiv a := \left.f'\right|_{x_i = x^*}\,, \notag \\
	b_{ij} &\equiv b := \left.\frac{\partial g}{\partial x_i}\right|_{x_i = x_j = x^*}\,, \notag \\
	c_{ij} &\equiv c := \left.\frac{\partial g}{\partial x_j}\right|_{x_i = x_j = x^*}\,.
\end{align}
We thus obtain
\begin{equation}
	\dot{\epsilon}_i = (a + bk_i)\epsilon_i + c\sum_{j = 1}^NA_{ij}\epsilon_j\,, \qquad i \in \{1,\ldots, N\}.
\end{equation}
If we assume that $g(x_i,x_j) = g(x_j)$, which is yet another major simplifying assumption (don't you love how many assumptions we're making?), we obtain
\begin{align}\label{simp2}
	\dot{x}_i &= f(x_i) + \sum_{j = 1}^N A_{ij}g(x_j)\,.
\end{align}
Consequently, $b = 0$ and	
\begin{align}\label{dyn3}	
	\dot{\bf \epsilon} &= (a{\bf I} + c {\bf A}){\bf \epsilon}\,,
\end{align}
where ${\bf I}$ is the $N \times N$ identity matrix.

An equilibrium of (\ref{dyn3}) is (locally) asymptotically stable if and only if all of the eigenvalues of ${\bf P} := a{\bf I} + c{\bf A} = {\bf P}^T$ are negative.  (The matrix ${\bf P}$ is symmetric, so all of its eigenvalues are guaranteed to be real.)  Let ${\bf w}_r$ denote an eigenvector of ${\bf A}$ with corresponding eigenvalue $\lambda_r$ \footnote{Note that we previously used the notation $\lambda$ in Sec.~\ref{SImodel} to represent the transmission rate in the SI model. In this section, we use it with appropriate subscript to represent the eigenvalues of ${\bf A}$.}.
It follows that
\begin{equation}
	(a{\bf I} + c{\bf A}){\bf w}_r = (a + c\lambda_r){\bf w}_r
\end{equation}
for all $r$ (where there are at most $N$ eigenvectors and there are guaranteed to be exactly $N$ of them if we are able to diagonalize ${\bf A}$), so ${\bf w}_r$ is \emph{also} an eigenvector of the matrix ${\bf P}$. Its corresponding eigenvalue for the matrix ${\bf P}$ is $(a + c\lambda _r)$.  For (local) asymptotic stability, we thus need $a + c\lambda_r < 0$ to hold for all $\lambda_r$.  This, in turn, implies that we need $a < 0$ because the adjacency matrix ${\bf A}$ is guaranteed to have both positive and negative eigenvalues \cite{newman2010}. We thus need (i) $\lambda_r < -a/c$ for $c > 0$ and (ii) $\lambda_r > -a/c$ for $c < 0$.  If (i) is satisfied for the most positive eigenvalue $\lambda_1$ of ${\bf A}$, then it (obviously) must be satisfied for all eigenvalues of ${\bf A}$.  If (ii) is satisfied for the most negative eigenvalue $\lambda_N$ of ${\bf A}$, then it (obviously) must be satisfied for all eigenvalues of ${\bf A}$.  It follows that
\begin{equation}
	\frac{1}{\lambda_N} < -\frac{c}{a} < \frac{1}{\lambda_1}\,,
\end{equation}	
which becomes much more insightful when we insert the definitions of $a$ and $c$.  This yields
\begin{align}\label{msc}
	\frac{1}{\lambda_N} < -\frac{\left.\frac{\partial g}{\partial x_j}\right|_{x_i = x_j = x^*}}{\left.f'\right|_{x = x^*}}< \frac{1}{\lambda_1}\,.
\end{align}
The left and right terms in equation (\ref{msc}), which is called a \emph{master stability condition}, depend \emph{only} on the structure of the network, and the central term depends \emph{only} on the nature (i.e., functional forms of the individual dynamics and of the coupling terms) of the dynamics.  In our opinion, that's really awesome!  Less enthusiastically but equally importantly, it also illustrates that the eigenvalues of adjacency matrices have important ramifications for dynamical behavior when studying dynamical systems on networks. Indeed, investigations of the spectra (i.e., set of eigenvalues) of adjacency matrices (and of other matrices, such as different types of graph Laplacians) can yield crucial insights about dynamical systems on networks \cite{newman2010,SoleRibalta2013Spectral}. Such insights have repeatedly been important in the analysis of dynamical systems on networks \cite{arenas-review,rom-review2014,piet-book}.

Now let's suppose that each node is associated to more than one variable.  We now write
\begin{align}\label{dyn4}
	\dot{{\bf x}}_i = {\bf f}_i({\bf x}_i) + \sum_{j = 1}^N A_{ij} {\bf g}_{ij}({\bf x}_i,{\bf x}_j)\,, \qquad i \in \{1, \ldots, N\}\,.
\end{align}
In other words, the variables and functions are now vectors.  As before, we do linear stability analysis, and we again derive an equation of the form
\begin{equation}
	\dot{{\bf \epsilon}} = {\bf M}{\bf \epsilon}\,,
\end{equation}	
where ${\bf \epsilon}$ and ${\bf M}$ are both $N \times N$ matrices.  The matrix component $\epsilon_{im}$ denotes the perturbation (in the linear stability analysis) of the $m$th variable on the $i$th node.  If we assume that the same vector function ${\bf f} \equiv {\bf f}_i$ describes the intrinsic dynamics on node $i$ and that the coupling function ${\bf g} \equiv {\bf g}_{ij}$ is the same for all pairs of nodes, then the components $M_{im,jn}$ of the matrix ${\bf M}$ [which we index using a double pair of indices $(i,m)$] are \cite{newman2010}
\begin{equation}
	M_{im,jn} = \delta_{ij}a_{mn} + \delta_{ij}k_ib_{mn} + A_{ij}c_{mn}\,,
\end{equation}
where
\begin{align} \label{gah}
	a_{mn} &:= \left.\frac{\partial f_m({\bf x})}{\partial x_n}\right|_{{\bf x} = {\bf x}^*}\,, \notag \\
	b_{mn} &:= \left.\frac{\partial g_m({\bf u,v})}{\partial u_n}\right|_{{\bf u}, {\bf v} = {\bf x}^*}\,, \notag \\
	c_{mn} &:= \left.\frac{\partial g_m({\bf u,v)}}{\partial v_n}\right|_{{\bf u}, {\bf v} = {\bf x}^*}\,,
\end{align}
and we are using the dummy variables ${\bf u} = {\bf x}_i$ and ${\bf v} = {\bf x}_j$ in equation (\ref{gah}) to prevent confusion.

Let's now assume once again that ${\bf g}({\bf x}_i,{\bf x}_j) = {\bf g}({\bf x}_j)$.  This yields
\begin{align}
	\dot{\epsilon}_{im} = \sum_{jn}M_{im,jn}{\epsilon}_{jn} =  \sum_{jn} \left(\delta_{ij}a_{mn} + A_{ij}c_{mn}\right){\epsilon}_{jn}\,.
\end{align}
As before, we assume that ${\bf M}$ is diagonalizable, and we expand ${\bf \epsilon}$ using the eigenvectors ${\bf v}_r$ of ${\bf M}$. We thus write
\begin{align}
	\epsilon_{im}(t) = \sum_{r}\alpha_{rm}(t)v_{ir}\,,
\end{align}
and we separately equate the coefficients of the independent eigenvectors ${\bf v}_r$ (with corresponding eigenvalues $\mu_r$).  The $i$th component of ${\bf v}_r$ is $v_{ir}$, and the coefficients $\alpha_{rm}(t)$ satisfy the matrix dynamical system
\begin{align}
	\dot{\bf \alpha}_r = \left({\bf a} + \lambda_r{\bf c}\right){\bf \alpha}_r(t)\,,
\end{align}
where $\lambda_r$ is the $r$th eigenvalue of the adjacency matrix.

Define $\sigma(\lambda)$ to be the largest positive real part among the eigenvalues of the matrix $\tilde{{\bf M}} := ({\bf a} + \lambda {\bf c})$.  The matrix $\tilde{{\bf M}}$ has size $\tilde{N} \times \tilde{N}$, where $\tilde{N}$ is the number of variables associated with each node.  For the dynamics to be (locally) asymptotically stable near the equilbrium point ${\bf x}^*$, we require that $\sigma(\lambda_r) < 0$ for all $r$. The function $\sigma(\lambda)$, which is an example of a \emph{master stability function} \cite{pecoracarroll1998}, tends to be easy to evaluate numerically. This is excellent news, because one can then use it readily to obtain interesting insights.

Now that we have slogged through the above calculations, let's review and illustrate what we can learn from it (and from its generalizations \cite{arenas-review}). The idea of studying an MSF (and MSC) is to have a general way to examine the effects of network structure on dynamical systems on networks.  As we have seen, the result of an MSF or MSC analysis is to derive a relation between the spectrum of a network's adjacency matrix (or some other matrix associated with a network) to the stability of some kind of qualitative behavior.  Obviously, it is of interest to see how this plays out on specific network architectures and for specific functions (and families of functions) that describe the dynamics of the individual oscillators (namely, $f_i$) and those that describe how oscillators interact with each other (namely, $g_{ij}$).

The use of MSFs and MSCs to investigate dynamical systems on networks is widespread and can be very insightful \cite{fink2000,nishikawa2003,qi2008,sun2011}. Reference~\cite{arenas-review} contains a lot of relevant discussion, although there does not appear (as far as we can tell) to be a dedicated review article on MSFs and MSCs. As discussed at length in \cite{arenas-review}, they have been employed to great effect in the examination of the stability of various types of dynamical behavior in coupled oscillators on networks, although there is far from a complete understanding of such phenomena. For instance, an MSF can give explicit conditions for how easily a dynamical system coupled via a network can attain a stable state (such as one that corresponds to synchrony of oscillators at the network's nodes), and it is useful to compare the relative ease of attaining stability in different families of networks. In some cases, it is possible to use an MSF to derive necessary conditions for the linear stability of a state in terms of the spectrum of a network's adjacency matrix (or some other matrix associated with it).  One can either compute the spectrum numerically or can take advantage of analytical expressions, approximate expressions, and bounds about the spectra of appropriate matrices. From such an analysis, it has been demonstrated that it tends to be very difficult in a ring network, in which each node is adjacent to its $2b$ nearest neighbors, to attain a synchronized stable state for a large class of oscillators and a large variety of ways to couple them \cite{arenas-review}. However, by adding a small number of ``shortcut'' edges (which connect distant nodes in the ring) to such a network \cite{smallworld-scholarpedia}, it is possible to significantly speed up the attainment of a stable state of synchronized oscillators for many types of coupled oscillators \cite{Barahona2002}. Importantly, because an MSF separates structural and dynamical properties, it is possible using MSFs to make broad statements about the stability of states in a large set of dynamical systems without having to individually explore every type of function $f_i$ and $g_{ij}$. Therein lies the power of the MSF approach. The primary weakness of an MSF approach is that it does not indicate the route that is taken towards a stable state, as it is concerned with linear stability \cite{arenas-review}.

Finally, we note that although our discussion has included numerous simplifying assumptions for expository convenience, many of them can be relaxed. More general situations (e.g., complex spectra from directed networks, bifurcation phenomena, etc.) necessitate more complicated expressions and analysis, but our discussion above nevertheless conveys the fundamental ideas of an MSF approach.  See \cite{arenas-review} for further discussion.

%%%%

\subsection{Other Approaches for Studying Dynamical Systems on Networks}

There are also other ways to relate network structure to dynamics.  In this section, we discuss a few of them very briefly.

As we discussed in Sec.~\ref{master}, using a master stability function can yield necessary and sufficient conditions for the linear stability of a state (e.g., a synchronized state) of a dynamical system on a network \cite{arenas-review}. There are also other approaches for determining conditions for the stability of a state of a dynamical system. For example, one way to obtain necessary conditions is to construct spanning trees of the Coates graph (i.e., the network with the self-edges removed) of the Jacobian matrix near that state \cite{lydo2012,lydo2013}.

Another way to examine the effects of network architecture on dynamics is through the investigation of \emph{coupled-cell networks} \cite{stewart2003}.  The structure of a coupled-cell network is a graph that indicates how cells are coupled and which cells are equivalent, and a ``multiarrow formalism" \cite{golub2005} allows pairs of nodes to have multiple types of connections between them.  This also allows one to examine dynamical systems on multilayer networks \cite{mikko-review,bocca-review}, and a tensorial formalism and the construction of ``quotient networks" \cite{stewart2003} can be very helpful for exploring general bifurcation phenomena and robust patterns on coupled-cell networks (see, e.g., \cite{golub2009}).

Isospectral compression, expansion, and other transformations can also be helpful for characterizing dynamical systems on networks \cite{webb-chaos2012,webb-nonlin2012}. Additionally, methods from algebraic geometry are also being used increasingly to elucidate qualitative behaviors (e.g., of the number and type the equilibria) for dynamical systems on networks \cite{craciun2005,craciun2006,algebra-kuramoto,badal2015}, and more recently for investigating such network behavior quantitatively (e.g., with data) using computational algebra and statistics
\cite{Harrington:2012us,maclean2015,gross2015}. Methods from computational topology are also being explored increasingly actively \cite{taylor2014}.

%%%%%

\subsection{Discrete-State Dynamics: Mean-Field Theories, Pair Approximations, and Higher-Order Approximations} \label{MFPA}

Several approaches to approximating global (i.e., network-scale) observables have been developed to try to understand the relationship between network structure and local (i.e., node-level) discrete-state dynamics. Given the local (stochastic, discrete-state) dynamics, the goal is to accurately predict some emergent characteristics of the dynamics (e.g., the number of nodes that are infected with a disease). If an approximation method is amenable to mathematical analysis, it can also be possible to use it to identify bifurcation points or critical parameters that affect the qualitative dynamics. In biological contagion models, for example, it is desirable to estimate the \emph{epidemic threshold} (i.e., the ratio of transmission rate to recovery rate) that, if equaled or exceeded, enables a disease to spread globally through a networked population.

Analytical approximation approaches vary in their complexity, and there is usually an associated trade-off in accuracy (as measured, for example, by comparing the prediction of the theory with a large-scale Monte Carlo simulation of the dynamics). Theories of mean-field (MF) type are most common, as they can provide reasonable---and in some cases, very high---levels of accuracy and are relatively straightforward to formulate.

We begin by introducing some typical MF approximation schemes, which we illustrate with an example biological contagion (namely, SI disease-spread dynamics) and a threshold model of a social contagion.

%%%%%

\subsubsection{Node-Based Approximation for the SI Model}\label{nodebased}

We begin by considering node-based approximation schemes for the SI model for biological epidemics (see Sec.~\ref{SImodel}).  We closely follow the presentation in Ref.~\cite{newman2010}. The simple dynamics allows one to clearly identify the important approximations in this example.

A \emph{node-based approximation} is one in which a variable $x_i$ is defined for every node $i$ in a network. In a given stochastic simulation of the system, $x_i$ takes the value $1$ when node $i$ is infected and the value $0$ when it is not. If one then considers an ensemble of stochastic simulations, the dynamics evolves differently in each realization, but can compute the (time-dependent) expectation $\left\langle x_i\right\rangle$ of $x_i$ over all simulations in the ensemble.  (In practice, one also needs to compare such an expectation to the sample mean over the simulations that one performs in practice.) To write an equation for the temporal evolution of $\left\langle x_i \right\rangle$, we first note that if $x_i$ is $0$, it can change to $1$ only when the disease is transmitted (at a rate $\lambda$) from an already-infected neighbor in the network. The ensemble-averaged quantities $\left\langle x_i\right\rangle$ thus obey the following set of differential equations:
\begin{equation}
	\frac{d \left\langle x_i\right\rangle}{d t } = \lambda \sum_j A_{i j}\left\langle (1-x_i)x_j\right\rangle\,, \quad i \in \{1,\ldots,N\}\,, \label{NBA}
\end{equation}
where $A_{i j}$ is the adjacency matrix (so the sum over $j$ gets nonzero contributions only from neighbors of node $i$) and the quantity $\left\langle (1-x_i)x_j\right\rangle$ is the probability (averaged over the ensemble of realizations) that node $i$ is susceptible and node $j$ is infected.

The set of equations (\ref{NBA}) is large---there is one equation for each node in the network---but a more serious issue is that it is not a closed system. To close equations (\ref{NBA}), we must either approximate the quantity $\left\langle (1-x_i)x_j\right\rangle$ in terms of the variables $\left\langle x_i\right\rangle$ or we need to derive an equation for its temporal evolution. For example, by assuming independence---i.e., no ``dynamical correlations'' \cite{Gleeson12} between the states of nodes $i$ and $j$, as we discuss in Sec.~\ref{MFdisc}---one can write
\begin{align}
	\langle (1-x_i) x_j \rangle &= \langle 1-x_i \rangle \langle x_j \rangle
=\left(1-\langle x_i \rangle\right) \langle x_j \rangle\,,\label{MFmap}
\end{align}
which allows one to solve for  $\langle x_i \rangle$.  We have just performed a \emph{moment closure} in which we have closed at the first moment to produce a \emph{mean-field theory}.

Alternatively, we can derive
\begin{align}\label{pair-si}
		\frac{d \langle s_i x_j \rangle}{dt} = -\lambda \langle s_i x_j \rangle + \lambda \sum_{k \neq i} A_{jk} \langle s_i s_j x_k\rangle - \lambda \sum_{l\neq j} A_{il} \langle x_l s_i x_j\rangle\,,
\end{align}		
where we have written $s_i=1-x_i$ for convenience.  However, we now have to either approximate the triplet terms (to obtain a so-called \emph{pair approximation} \cite{newman2010}, once we also express the ${\langle s_i s_j \rangle}$ and ${\langle x_i x_j \rangle}$ pair terms in terms of $\langle s_i x_j \rangle$) or derive dynamical equations for the triplet terms. Such equations will include quadruplet terms, of course.

As an explicit example of closing equations (\ref{pair-si}) to obtain a pair approximation \cite{newman2010}, one can use Bayes theorem to derive the approximations
\begin{equation*}
	  \langle s_is_jx_k \rangle \approx \frac{\langle s_is_j\rangle \langle s_jx_k\rangle}{\langle s_j\rangle} \quad \text{ and } \quad \langle x_l s_i x_j \rangle \approx \frac{\langle x_ls_i \rangle \langle s_ix_j\rangle}{\langle s_i\rangle}\,.
\end{equation*}
We also use	
\begin{equation*}
	\langle s_i s_j \rangle = \langle s_i(1-x_j)\rangle = \langle s_i \rangle - \langle s_i x_j \rangle \,,
\end{equation*}	
and we thereby derive the closed system of equations
\begin{align}\label{pair}
	\frac{d \langle s_i x_j \rangle}{dt} &= -\lambda \langle s_i x_j \rangle
		+ \lambda \sum_{k\neq i} A_{jk} \frac{(\langle s_i \rangle - \langle s_i x_j \rangle) \langle s_jx_k\rangle}{\langle s_j\rangle}\
		-\lambda \sum_{l \neq j} A_{il} \frac{\langle x_ls_i \rangle \langle s_ix_j\rangle}{\langle s_i\rangle}	 \notag \\
		&= -\lambda \langle s_i x_j \rangle
		+ \lambda \frac{\langle s_i \rangle - \langle s_i x_j \rangle}{\langle s_j\rangle}\sum_{k\neq i} A_{jk}\langle s_jx_k\rangle\
		-\frac{\lambda}{\langle s_i\rangle}\sum_{l \neq j} A_{il}\langle x_ls_i \rangle \langle s_ix_j\rangle \,,
\end{align}
which, along with equation (\ref{NBA}) and the expression $x_i  = 1 - s_i$, constitutes a pair approximation.

The  above discussion has given some examples of the moment-closure problems that often arise for stochastic dynamics on networks.  See Ref.~\cite{miller-tute2014} for much more detail on moment closure, and see Ref.~\cite{house2014} for a discussion of the use of algebraic methods for moment closure as an alternative to more common approaches like what we have just discussed.

%%%%%

\subsubsection{Degree-Based MF Approximation for the SI Model}\label{degreebased}

Dealing with the issue of moment closure requires truncating a hierarchy of differential equations at some stage. Given the complexity of the equations that arise, it is common to reduce the number of equations by assuming that all nodes of degree $k$ behave in a manner that is dynamically similar. In applying such a scheme, which is called a \emph{degree-based approximation} in Ref.~\cite{newman2010}, one is making the assumption that it is reasonable to consider the dynamics (at least as concerns the observables of interest) on a configuration-model network.

Suppose that node $i$ has degree $k$. We then replace $\left\langle  x_i\right \rangle$ in equation~(\ref{NBA}) with a new variable $\rho_k(t)$, which is defined as the fraction of nodes of degree $k$ that are infected at time $t$. With the MF approximation (\ref{MFmap}), the right-hand side of equation~(\ref{NBA}) becomes
\begin{equation}
	\lambda \left(1-\left\langle x_i\right\rangle\right)\sum_j A_{i j}\left\langle x_j\right\rangle \,.
\end{equation}
In the degree-based approximation, we replace $1-\left\langle x_i\right\rangle$ by $1-\rho_k$, and we replace the sum over neighbors by $k\, \omega$, where
\begin{equation}
	\omega(t) = \sum_k \frac{k}{z} P_k \rho_k \label{defw}
\end{equation}
is the probability that a node at the end of a randomly-chosen edge is infected.  Specifically, we choose this edge uniformly at random from all edges of node $i$, and our use of $P_k$ in equation (\ref{defw}) indicates that we are assuming a configuration-model network. Consequently, the MF degree-based approximation for the SI model is the system of equations
\begin{equation}
	\frac{d \rho_k}{d t} = \lambda k (1-\rho_k) \omega\,, \label{MFmap2}
\end{equation}
where $\omega$ is given by equation~(\ref{defw}). Note that the system (\ref{MFmap2}) contains one equation for each degree class $k$ in the network, and typically this number is much smaller than the number $N$ of nodes, so the dimension of the system (\ref{MFmap2}) is considerably lower than that of (\ref{NBA}). To take an extreme example, consider a $z$-regular network, where every node has exactly $z$ neighbors. In this case, equation~(\ref{MFmap2}) reduces to a single equation,
\begin{equation}
	\frac{d \rho}{d t} = \lambda z (1-\rho)\rho\,, \label{MFzreg}
\end{equation}
for the the fraction $\rho(t)$ of infected nodes. This is the well-known (and analytically solvable) logistic differential equation that appears in SI models of homogeneous, well-mixed populations \cite{ccc}.

In the following sections, we will concentrate on degree-based approximations for other binary-state dynamics. In the Appendix, we will consider a hierarchy of approximation schemes that yield successively better accuracy.

%%%%%

\subsubsection{Degree-Based MF Approximation for a Threshold Model}\label{mf-thresh}

We now derive a MF approximation for a threshold model of social contagion. (See Sec.~\ref{thresh} for a discussion of such process.) In this section, we give an \emph{ad hoc} derivation that highlights some of the important assumptions of MF approximations.  In the Appendix, we detail a more systematic approach for deriving MF (and other, higher-accuracy) approximations.

\begin{figure}
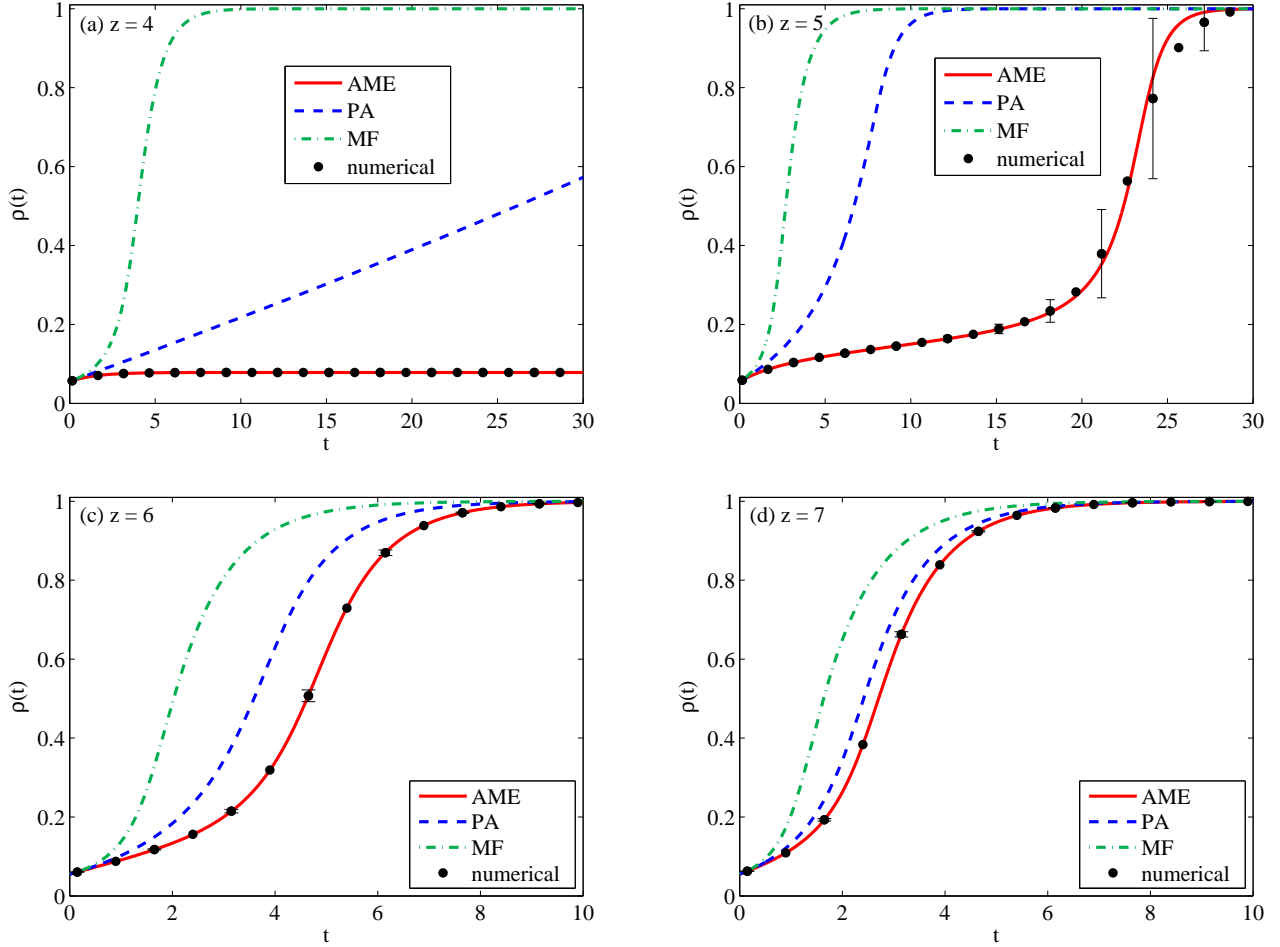

\centering
\epsfig{figure=fig_draw_thresh_Frontiers_z4.eps,width=8.8cm}
\epsfig{figure=fig_draw_thresh_Frontiers_z5.eps,width=8.8cm}
\epsfig{figure=fig_draw_thresh_Frontiers_z6.eps,width=8.8cm}
\epsfig{figure=fig_draw_thresh_Frontiers_z7.eps,width=8.8cm}
\caption{Fraction $\rho(t)$ of infected nodes in a Watts threshold model on  $z$-regular random graphs (i.e., $P_k=\delta_{k,z}$) for $z \in \{4,5,6,7\}$. We show the results of the mean-field (MF) approximation (\ref{MFeqn}) using the green dash-dotted curves, and also show the results of pair approximation (PA) and approximate master equation (AME) schemes. (We discuss these approximations in the Appendix.) The initial fraction of infected nodes is $\rho(0) = 0.055$ in all cases, and all nodes in each network have the same threshold level of $R = 2/z$ (so that a node activates if $2$ or more of its neighbors are active). The results of numerical simulations (see Sec.~\ref{numericalmethods}) are for networks with $N=10^5$ nodes, and they use a time step of $dt=10^{-5}$. The black symbols show the means over 24 realizations, and the error bars indicate one standard deviation above and below the mean.
}\label{fig1}
\end{figure}

Let's consider a Watts threshold model with asynchronous updating. For simplicity, as in Eq.~(\ref{MFzreg}), we assume that the network is $z$-regular, but one can readily generalize the derivation to networks with arbitrary degree distributions and degree-degree correlations \cite{PastorSatorras01,Boguna02}. Define $\rho(t)$ to be the fraction of nodes that are active at time $t$; we assume that a given seed fraction $\rho(0)$ of nodes (which we choose uniformly at random) are initially activated. To derive a MF approximation, we consider how $\rho(t)$ changes in time. Consider an updating event, in which we have randomly selected a node for a possible change of state; the probability that the chosen node is inactive at time $t$ is $1-\rho(t)$. We want to calculate the probability that $m$ of the neighbors of the node are active, and we then compare the fraction $m/z$ with the threshold $R$ of the node. The probability that $m/z \geq R$ is the probability that the updating node becomes active, and the activation of a node increases $\rho$. To continue, we will need to make two assumptions.

In the first of the assumptions, we suppose that all of the neighbors of the selected node are (independently) active with probability $\rho$(t). This independence assumption is an important one; as we will discuss in assumption (1) of Sec.~\ref{MFdisc}), this assumption cannot be exactly true on networks that contain triangles or other short cycles. However, with this assumption, we can write the probability that the chosen node has $m$ active neighbors as the binomial distribution
\begin{equation}
	B_{z,m}\left(\rho\right) = \left(\begin{array}{c} z \\m\end{array}\right) \rho^m(1-\rho)^{z-m}\,,
\end{equation}
because we are considering $z$ neighbors, who are each (independently) active with a probability of $\rho(t)$.

The second important MF assumption arises when we suppose that the probability that the updating node is inactive \emph{and} that it has $m$ active neighbors is given by $\left[1-\rho(t)\right)B_{z,m}\left(\rho(t)\right]$. To obtain this expression, we have multiplied the probability $1-\rho$ of the node being inactive by the probability $B_{z,m}(\rho)$ that it has $m$ active neighbors. Calculating a joint probability in this way is justified if the two events---in this case, that the ``updating node is inactive'' and that the ``updating node has $m$ active neighbors''---are independent. Consequently, the second important MF assumption (see assumption (2) in our discussion in Sec.~\ref{MFdisc}) is that the state (active or inactive) of an updating node is independent of the states of its neighbors. This assumption cannot be exactly correct; for instance, a node whose neighbors are all active is more likely to be active than a node with no active neighbors. The dynamics of the system cause correlations between the states of nodes and their neighbors, but our MF assumptions are ignoring such ``dynamical correlations.''

Returning to our derivation, we have thus far agreed---we didn't really give you a choice, to be honest, except possibly to look at the more complicated approach in the Appendix---to approximate the probability that the updating node is inactive and has $m$ active neighbors by $\left[1-\rho(t)\right)B_{z,m}\left(\rho(t)\right]$. We now ask the following question: What is the probability that such a node will become active and thereby increase the active fraction $\rho(t)$? According to the update rule in the Watts model---which becomes equivalent to the Centola-Macy update rule on a $z$-regular network---the node will become active if the fraction $m/z$ of active neighbors equals or exceeds its threshold $R$.
Recall that we chose $R$ from a predefined distribution, so we need to determine the probability that a threshold value $R$ drawn from this distribution is less than or equal to the value $m/z$.  This probability is given by
\begin{equation}
	\text{Prob}\left(R\le \frac{m}{z}\right) = C\left(\frac{m}{z}\right)\,,
\end{equation}
where $C$ is the cumulative distribution function (CDF) of the thresholds.

Putting together the probabilities that we calculated above and summing over the possible values of $m$, we obtain the following MF approximation for the fraction $\rho(t)$ of active nodes:
\begin{equation}
	\frac{d \rho}{d t} = \left( 1- \rho\right) \sum_{m=0}^z B_{z,m}(\rho) C\left(\frac{m}{z}\right)\,. \label{MFeqn}
\end{equation}
Equation (\ref{MFeqn}) is a nonlinear ODE for $\rho(t)$, with an initial condition of $\rho(0)$, and one can easily solve it numerically to obtain the time-dependent MF prediction for $\rho(t)$. In Fig.~\ref{fig1}, we show a few examples, in which we compare the solutions of the MF equation (green dash-dotted curves) with ensemble averages of direct (stochastic) numerical simulations of the Watts model (black symbols).  See Sec.~\ref{MCsim} for a description of such numerical simulations. Although the MF predictions are qualitatively correct for the better-connected networks ($z=6$ and $z=7$), the quantitative agreement tends to be poor. Moreover, on low-degree networks (e.g., $z=4$), the MF approximation is even qualitatively incorrect: it predicts that the social contagion spreads through the whole network, whereas in fact only a very small fraction of nodes ever adopt the contagion. In these cases, one can use higher-accuracy approximations (as we describe in the Appendix), albeit at the cost of increased complexity of the dynamical system and of the derivation. 

%%%%%%

\subsubsection{Discussion of MF Approximation for Discrete-State Dynamics} \label{MFdisc}

As we have seen in our discussions in Sec.~\ref{MFPA}, the main assumptions that underly the derivation of MF theories are as follows \cite{Gleeson12}:
\begin{enumerate}
\item[(1)] Absence of local clustering: When the state of node $i$ is being updated, the states of the neighbors of node $i$ are considered to be independent of each other. This holds, for example, when one makes a \emph{locally tree-like} assumption on the structure of a network. With such a structural assumption, the cycles in a network become negligible as $N \rightarrow \infty$, so there are very few pairs of neighbors to consider in the first place. (This motivates the terse name for the assumption.)
\item[(2)] Absence of \emph{dynamical correlations}: When updating the state of node $i$, its own state and the state of its neighbors are assumed to be independent. Such dynamical correlations are rather different from \emph{structural correlations} like degree-degree correlations. Note that dynamical correlations can have strong effects even in $z$-regular random graphs (see Sec.~\ref{mf-thresh} and Fig.~\ref{fig1}), whereas degree-degree correlations play no role in this situation by construction \cite{Gleeson11}.
\item[(3)] Absence of modularity: It is usually assumed that one can describe the state of every node of given degree $k$ using a single quantity (namely, by the mean over the class of degree-$k$ nodes). However, this may not be the case if a network has significant modularity, as degree-$k$ nodes can often be located in different communities, and there could also be other sources of diversity among the degree-$k$ nodes \footnote{It may also be useful to develop analogous approximations that use structural characteristics other than degree.}.
\end{enumerate}

One can capture dynamical correlations in part by using generalizations of MF theories that incorporate information on the joint distribution of node states at the ends of a random edge in a network. Such theories are often called \emph{pair approximations} (PA) and are more complicated to derive than MF theories \cite{miller-tute2014}. However, they also tend to be more accurate than MF theories.  One can also include the dynamics of triplets, quadruplets, and as many nodes as one wants by considering more general \emph{motif expansions} \cite{thilotute,kiss2013,miller-tute2014}.

It is also possible to achieve very accurate approximations by considering compartmental models in which one considers the states of all neighbors when updating a node. Such approaches are expensive computationally, but it has been demonstrated that they can yield high accuracy for models of biological contagions \cite{Marceau10,Lindquist11}. They have also been successfully generalized for a wide class of binary-state models \cite{gleeson13}. Reference \cite{gleeson13} also illustrates how to systematically derive PA and MF theories from higher-order compartmental models. See the Appendix.

Despite the fact that MF assumptions (1)--(3) are violated for many (and perhaps most?) real-world networks, MF theories often give reasonably accurate predictions of global dynamics---at least for well-connected networks and when the dynamical system under consideration is far from any bifurcation points. However, for sparse networks and for accurate bifurcation analysis, one often needs to work harder to achieve higher levels of accuracy. This motivates a move beyond pair approximations to methods (such as motif expansions) that include more information on the states of neighbors of each node (see the Appendix) \cite{thilotute,BohmeGrossPRE11}.

MF theories and their generalizations provide analytical approaches for reducing the dimensionality of a dynamical system on a network via an approximation scheme.  Alternatively, one can take a philosophically similar approach (for both dynamics on networks and dynamics of networks) using only computations.  See Refs.~\cite{yannis2,yannis1} for examples.

%%%%%

\subsection{Additional Considerations}

As  discussed in the Appendix, there exist rather sophisticated approximation methods for binary-state dynamics on configuration-model networks. Moreover, for monotonic binary dynamics, there has also been significant progress in extending theories to networks with degree-degree correlations, clustering, modularity, and multilayer structures (see Sec.~\ref{deriveMFPA}). Indeed, degree-degree correlations have been incorporated into many MF and PA methods, giving approaches that are sometimes called \emph{heterogeneous-MF} and \emph{heterogeneous-PA} theories.

However, despite this progress, much remains to be done. Moving beyond binary-state dynamics is a significant challenge. There has been notable progress in some dynamical systems that include three states---particularly when the dynamics remains monotonic (i.e., transitions between states still occur only in one direction). Examples include the multi-stage complex contagion model of social influence in Ref.~\cite{MelnikChaos13} and high-accuracy approximations for SIR disease spread \cite{Lindquist11,Miller}. It would be very interesting---though also rather challenging---to extend the compartmental AME approach to dynamics with more than two states and to subsequently use the results of such analysis to derive PA and MF approximations like (\ref{MF}) and (\ref{PA}) in a systematic fashion. Other open problems include developing high-accuracy approximations of nodal dynamics with a continuum of states---e.g., with differential equations at each node, as in the Kuramoto oscillator model \cite{ichi2004}---and understanding the novel features observed in non-monotonic binary dynamics with synchronous updates in \cite{DoddsDeckerPRL}.

%%%%%%%

\section{Software Implementation} \label{numericalmethods}

In this section, we briefly discuss practical issues for simulating dynamical systems on networks.

%%%%%%

\subsection{Stochastic Simulations (i.e., Monte Carlo Simulation)}\label{MCsim}

As we noted in Sec.~\ref{stochasticbinarystate}, it is straightforward to implement a Monte Carlo simulation of stochastic discrete-state dynamics. The special case of monotonic threshold dynamics is especially simple, so we describe it briefly in the context of the Watts model. (See Ref.~\cite{holme-tute2014} for discussions of fast algorithms for epidemic models on networks.)

Given an adjacency matrix $\mathbf{A}$ and an $N\times 1$ vector $\mathbf{v}$ that stores the states of each node (at a given time, $v_i=0$ if node $i$ is susceptible and $v_i=1$ if node $i$ is infected), we calculate the number of infected neighbors $m_i$ of node $i$ using matrix multiplication:
\begin{equation}
	\mathbf{m}=\mathbf{A} \mathbf{v}\,. \label{mv}
\end{equation}
Similarly, the degree $k_i$ of each node $i$ is the $i$th element of the vector $\mathbf{k}$ defined by
\begin{equation}
	\mathbf{k} = \mathbf{A} \mathbf{1}\,,
\end{equation}
where $\mathbf{1}$ represents the $N\times 1$ vector $(1,1,1,\ldots,1)^T$. At each time step in an asynchronous updating scheme (see Sec.~\ref{async}), a node $i$ is chosen uniformly at random for updating, and $v_i$ is set to 1 if $m_i/k_i$ equals or exceeds the threshold $r_i$ of node $i$.

One then uses the updated state vector $\mathbf{v}$ to calculate the updated $\mathbf{m}$ vector from equation (\ref{mv}), and the simulation continues to the next time step. One can terminate the temporal evolution when the condition
\begin{equation}
	\frac{m_i}{k_i}<r_i \quad \text{ for all nodes }\quad i \quad \text{ with } \quad v_i=0
\end{equation}
is satisfied, as this implies that no further susceptible nodes can be infected and thus that the system has reached a steady state. One can show \cite{Gleeson07b} that the steady state of monotonic threshold dynamics is independent of whether one employs asynchronous or synchronous updating, so one can accelerate the algorithm described above by updating all nodes simultaneously in each time step if one is interested only in the steady state.  (Naturally, it is often the case that one wishes to explore phenomena other than a steady state \cite{MelnikChaos13}.)

\subsection{Differential Equation Solvers for Theories}

{\sc Octave}/{\sc Matlab} codes for implementing and solving the systems of ODEs that arise from various approximation schemes for stochastic binary-state dynamics are available at \url{www.ul.ie/gleesonj/solve_AME}. In these codes, a user specifies as inputs the degree distribution $P_k$ of a network, the infection rate $F_{k,m}$ and recovery rate $R_{k,m}$ of the dynamics, and the initial fraction $\rho(0)$ of infected nodes. The code then automatically implements the AME equations (see Ref.~\cite{gleeson13} and the discussion in the Appendix), the PA equations (\ref{PA}), and the MF equations (\ref{MF}) as systems of ODEs and then solves them using standard numerical techniques. Results are output as plots that show the infected fraction $\rho(t)$ of nodes and the fraction of edges that connect susceptible nodes to infected nodes as function of time.

Additionally, Gerd Zschaler and Thilo Gross have posted software \cite{thilosoftware} for simulating dynamical systems on time-dependent networks at \url{http://www.biond.org/node/352}.  See Sec.~\ref{temporal} for a brief discussion of such systems.

%%%%%

\section{Dynamical Systems on Dynamical Networks} \label{temporal}

The study of dynamical systems on time-dependent networks has become extremely popular recently, but there are also much older quantitative studies of such phenomena.  For example, Farmer et al. \cite{farmer1986} and Bagley et al. \cite{bagley1989} used such a framework more than two decades ago in studies of chemical reactions.

Our discussion in prior sections focused on dynamics that occur on time-independent networks and on ensembles of such networks.  If we are considering a single time-independent (i.e., ``static'') network, we are assuming that the network's structure does not change on the time scale of the nodal dynamics or at least that it changes so little that it is permissible to pretend that it is time-independent \footnote{In the physics literature, such situations with extremely slow structural dynamics are sometimes called ``quenched,'' because the networks are almost frozen.}. If one is studying a dynamical process on an ensemble of time-independent random graphs, there are two primary possibilities:
\begin{itemize}
\item{We are in the same situation as above as concerns the balance of time scales (so we have an ensemble of static networks), and the randomness is employed because of uncertainties about the network structure (e.g., perhaps one believes in a certain situation that only degree distribution is important, such that one fixes the degree distribution and randomizes everything else).}
\item{The network structure changes on such a fast time scale that it is only reasonable to use a random-graph ensemble to describe its properties. That is, instead of an ensemble of ``static'' networks, we need to use a description as a statistically stationary probability distribution \footnote{In the physics literature, this situation is sometimes called ``annealed.''}, as only appropriate averaged properties of the network are reliable.
}
\end{itemize}

There are many situations in which a network itself can be dynamic, with edges (or edge weights) and/or nodes changing in time, perhaps in response to a dynamical system running on the nodes.  In general, three basic situations are possible \cite{vespPRL2012,rom-review2014}:
\begin{itemize}
\item{A dynamical system on a network runs on a much faster time scale than the dynamics of the network itself.  In this case, it is reasonable (at least to a first approximation) to assume that a network is time-independent.}
\item{The dynamics of a network are much faster than the dynamical system that one is examining on that network.  In this case, it is reasonable (at least to a first approximation) to assume that the states of the network nodes are fixed (e.g., a ``susceptible" node will not change to being ``infected") and to consider only the dynamics of the underlying network.  (One can also make a similar comment about the states of edges if that is what one is studying.) However, if the dynamics of the network structure are too fast, then it may be necessary to consider ensembles of time-independent networks because only the measurement of suitable averaged properties is appropriate.
}
\item{The dynamics on a network and the dynamics of the network itself operate on comparable time scales, so it is not reasonable to ignore either of them.  Such networks are sometimes called ``adaptive networks" \cite{thilo-adaptive,thilo-adapt2}.}
%\item{One other balance of the time scales also leads to the examination of dynamical systems on static networks. Suppose that network structure changes so rapidly that only some average properties of a network are reliable.  One can then posit that only ensemble averages are appropriate---where the particular ensemble is determined from the properties in question---so that one can study a dynamical system on an ensemble of static networks that are constructed using an appropriate random-graph model. (One can also imagine a similar situation with node states that change so rapidly, such that one examines time-dependent network structure but with node and/or edge states that are average values from some distribution.)}
\end{itemize}

The dynamics of networks can have a profound impact on dynamical systems on networks, and this has now been explored in numerous papers. See, e.g., \cite{holme12,holme13,vespPRL2012,belykh2004,bollt-moving2004,petter-plosone,petter-physicaA,mikko-slow,masuda-slow,till2012,till-chapter,vesp-actdriven,colizza-prx2015,kaneko2002,pfitz2013,ingo2013,horvath2014,volzmeyers2007,millervolz2013} and many other references.)  The bursty---and, more generally, non-Poisson---nature of interactions in temporal networks affects not only the effective weights of static interactions derived from averaging over such interactions but also the behavioral of dynamical processes on time-dependent networks versus those on static networks obtained from averaging \cite{till2012,till-chapter,vespPRL2012,boguna2013}. It is also important to develop computational techniques, such as Gillespie algorithms, for dynamical processes on temporal networks \cite{vestergaard2015}.

A good example of a dynamical system on a time-dependent (or ``temporal" \cite{holme12,holme13}) network is an adaptive SIS model \cite{Gross06}, in which a susceptible node can break any edge it has to an infected node and rewire to a randomly-chosen susceptible node. Compartmental approaches of the AME type have been applied successfully to this model \cite{Marceau10}, and they agree very well with the results of numerical simulation. Similar approaches were also used to study a two-opinion adaptive voter model in Ref.~\cite{Durrett12}, which built on the model in \cite{holme2006}.  (This model was generalized to $n > 2$ opinions in \cite{billshi2013} and to include mechanisms that reinforce local clustering in Ref.~\cite{nishant2013}.) In Ref.~\cite{Durrett12}, an edge of a network is chosen uniformly at random at each time step. If the opinions of the two nodes attached to the edge are different, then one node imitates the opinion of the other with a probability of $1-\alpha$.  However, with probability $\alpha$, the edge is broken and one node instead makes a new connection to another node (chosen, depending on the variant of the model, either at random from the whole network, or from those who hold the same opinion as the node). On a finite network, such adaptive voter dynamics eventually yields a steady state in which a network can contain disconnected components, where each component contains only nodes who share the same opinion. Adaptive voter dynamics has been a particularly challenging test for approximation methods. The review \cite{thilotute} carefully examined such a model and concluded that none of the approximation schemes that have been developed are able to give fully satisfactory results in all regions of parameter space.

A nice way to examine dynamical systems on temporal networks is using so-called \emph{activity-driven networks} \cite{vesp-actdriven}.  One constructs an ``activity potential" for each agent in such a network to encapsulate the number of interactions that he/she performs in a characteristic time window.  One thereby attempts to characterize interactions between agents, and activity rates can come either from specified functions or from empirical data. Using the activity potentials (which can be different for different agents), one can construct an instantaneous temporal description of network dynamics.  This approach has led to insights on phenomena such as the emergence of strong ties in temporal communication networks \cite{karsai-vesp1} and how to control contagions in temporal networks \cite{karsai-vesp2}.

It can also be useful and interesting to take a statistical perspective on dynamical systems on temporal networks (see, e.g., \cite{snijders2001,snijders2010}). For example, one can use a time-dependent generalization of exponential random graph models (ERGMs) \cite{robins2013} and thereby study a temporal ERGM (TERGM) \cite{tergm-skyler}.

When studying dynamical systems on time-dependent networks, there is a lively debate about whether dynamics of networks slow down dynamical processes or speed them up \cite{mikko-slow,masuda-slow,petter-plosone,petter-physicaA,holme12,holme13,pfitz2013,ingo2013} (and, naturally, the answer is different for different dynamical systems), and it is desirable to examine such effects on a broad variety of dynamical systems.  For example, bursty communication patterns can either speed up or slow down adoption speed in threshold models, and temporal-network generalizations of the Watts (i.e., fractional threshold) and Centola-Macy (i.e., absolute threshold) complex contagion models exhibit interesting differences \cite{petter-plosone,petter-physicaA}.

%%%%%

\section{Other Resources} \label{sec:OtherResources}

We now list several references that can complement the present article and indicate the particular directions for which we think you will find them helpful.
\begin{itemize}

\item Reference \cite{stro01} is a friendly introduction to dynamical systems on networks.  It gives the state of play in 2001, so it is out of date in many respects.  (Network science is a young, immature field that advances quickly.)

\item Reference \cite{motter-expo} is a more recent (but very short) expository article about dynamical systems on networks.

\item Reference \cite{vesp2012-review} is a recent review article that covers dynamical systems on networks.  It is not very technical, but it provides a big-picture overview of several areas.

\item Reference \cite{newman2010} is a very good textbook on network science in general.  Some of its later chapters cover dynamical systems on networks.  For example, some of our discussion of general considerations in Sec.~\ref{general} follow part of the presentation in this book, which also has thorough discussions of processes like percolation and biological contagions. It also discusses the use of mean-field theories and related techniques for studying such dynamical processes, and we drew on some of those discussions for portions of our presentation. In particular, in Sec.~\ref{nodebased}, we largely followed the presentation in \cite{newman2010}.

\item Reference \cite{kolac2009} is a textbook on networks that takes a statistical perspective.  It includes material on dynamical systems on networks.

\item Reference \cite{barratbook} is a book devoted to dynamical systems on networks from a physics perspective.

\item Reference \cite{rom-review2014} is a recent, extensive review of epidemic processes on networks.

\item Reference \cite{miller-tute2014} is a review on epidemic spreading in networks that focuses on moment-closure methods and discusses the challenges in generalizing methods from configuration-model networks to more complicated random-graph ensembles.

\item Reference \cite{thilotute} is a tutorial article that discusses moment-closure approximations for discrete adaptive networks, in which a discrete dynamical system is coupled to a network that changes in discrete time.  Reference \cite{thilo-adaptive} reviews research on adaptive networks.

\item Reference \cite{loreto2009} is a review of social dynamics from the perspective of statistical physics.

\item Reference \cite{jackson2014} is a survey concerning games on networks.

\item References \cite{mikko-review,bocca-review,salehi2014,goh-review,perc-multilayer} all include extensive discussions that review work on various types of dynamical processes on multilayer networks.  The body of such work is exploding rapidly, and these four articles offer four different perspectives on such topics.

\item Reference \cite{dorog2008} is a review article that discusses analytical methods to study critical phenomena on networks.  It is more difficult to digest than the above articles, but it has nice discussions of phase transitions, percolation, synchronization, and many other phenomena.

\end{itemize}

%%%%

\section{Conclusion, Outlook, and Open Problems} \label{conclude}

In this paper, we have given a tutorial for studying dynamical systems on networks.  By reading this paper, you should now have a reasonable understanding of (1) why it is interesting and desirable to study dynamical systems on networks; (2) several of the popular families of problems and models; (3) basic considerations about dynamical systems on networks; (4) the use and range of validity of techniques for analytical approximation such as mean-field theories, pair approximations, and higher-order motif expansions; (5) software implementation of both direct numerical simulations and for simulating equations that result from the aforementioned approximation methods; and (6) time-scale issues and challenges for investigating dynamical systems on time-dependent networks.  As this paper is a tutorial rather than a literature review, there is a lot that we haven't covered, and we strongly encourage you to scour the literature for interesting problems to study and generalize.  The primary purpose of this paper is to equip you with the background knowledge to be able to do so successfully. We have also provided numerous references to get you started.

Before saying ``goodbye," it is also worth commenting on some of the particularly challenging problems that are available.  For example, although one can study problems by purely computational means, it is desirable (when possible) to try to develop analytical techniques in order to gain insights on these problems.  In network science, mathematically rigorous results tend to be rare---see, e.g., \cite{durrettbook2007,durrettpnas2010}---but approximations and heuristic techniques have been employed on many toy problems.  A key goal is to find the ``next easiest" sets of dynamical systems on networks to study using more general versions of these techniques and to use them to help develop these methods further.  Many of the systems that have been investigated in the literature are either some kind of percolation or fairly (or even very) closely related to percolation, and it is important to move beyond that.

There are many open issues in the study of dynamical systems on time-dependent networks. One needs to examine the balance of time scales for dynamics on networks versus the dynamics of the networks themselves \cite{vespPRL2012}, and the effects that this has on the validity and choices of analytical techniques to use is also very important.  Most of the studied situations with both dynamics on networks and dynamics of networks tend to be rather unrealistic, so this is an area that is particularly ripe for further study.

Finally, many networks are \emph{multiplex} (i.e., include multiple types of edges) or have other \emph{multilayer} features \cite{mikko-review,bocca-review}.  The existence of multiple layers over which dynamics can occur and the possibility of both structural and dynamical correlations between layers offers another rich set of opportunities to study dynamical systems on networks.  The investigation of dynamical systems on multilayer networks is only in its infancy, and this area is also loaded with a rich set of problems \cite{mikko-review,bocca-review,salehi2014,goh-review}.

Goodbye.

%%%%%%%

\section*{Acknowledgements}

We were both supported by the European Commission FET-Proactive project PLEXMATH (Grant No. 317614).  MAP also acknowledges a grant (EP/J001759/1) from the EPSRC, and JPG acknowledges funding from Science Foundation Ireland (Grants No.~11/PI/1026, 12/IA/1683, 09/SRC/E1780). We acknowledge the SFI/HEA Irish Centre for High-End Computing (ICHEC) for the provision of computational facilities. We thank Alex Arenas, Vittoria Colizza, Marty Golubitsky, Heather Harrington, Matt Jackson, Vincent Jansen, Zoe Kelly, Barbara Mahler, Joel Miller, Yamir Moreno, Edward Rolls, Stylianos Scarlatos, Ingo Scholtes, Bernadette Stolz, and two anonymous referees for helpful comments.

%%%%%%

%\bibliography{frontiers3,frontiers-JG3}
%\bibliographystyle{abbrv}

%%%%

%%%%%%

\appendix
\section*{Appendix: High-Accuracy Approximation Methods for General Binary-State Dynamics}

%%%%%%

\subsection{High-Accuracy Approximations for Binary-State Dynamics}\label{AMEsec}

To illustrate some general concepts for how network structure affects dynamics, let's examine a class of \emph{stochastic binary-state dynamics} on configuration-model networks.  Recall from Sec.~\ref{basic} that a network in a \emph{configuration-model} ensemble is specified by a degree distribution $P_k$ but is otherwise maximally random \cite{bollobas1998,newman2010}: pairs of stubs (i.e., ends of edges) are connected to each other uniformly at random, so there are no correlations in the generative model, and taking the $N \rightarrow \infty$ limit guarantees negligible clustering (as long as the first and second moments increase sufficiently slowly in the limit).
In this section, we introduce a general class of binary-state dynamics and then examine approximation methods at mean-field, pair-approximation, and higher-order levels \cite{miller-tute2014}. We choose to focus on approximations that yield deterministic ODE systems. Deterministic approximations are valid when the initial seed size of, for example, infected nodes is sufficiently large so that one can neglect stochastic (i.e., realization-to-realization) fluctuations. Otherwise, it is necessary to use methods based on branching processes \cite{Noel09,Noel12,Noel13,Gleeson13b,Gleeson15}.

%%%%%%%%

\subsubsection{Stochastic Binary-State Dynamics}\label{stochasticbinarystate}

In a binary-state dynamical system, each node is in one of two states at any time. For convenience, we will usually refer to these states as \emph{susceptible} and \emph{infected} in our discussion.  Recall, however, that such states could actually mean ``spin up'' versus ``spin down'' magnetic dipoles in an Ising spin model, ``inactive" versus ``active" nodes in a social system, and so on.  We discussed many examples of binary-state dynamics in Sec.~\ref{examples}.  See, e.g., our discussions of percolation models \ref{percolation}, biological contagions \ref{simplecontagions}, and social contagions \ref{influence}. Additionally, we will consider asynchronous updating and local rules for node updates; in other words, the transition rates depend only on the state of a node and on the states of its immediate neighbors in the network.  Consequently, an updating node that is susceptible becomes infected with probability $F_{k,m}\, dt$, where $k$ is the node's degree and $m$ is the number of its neighbors that are infected. We refer to the function $F_{k,m}$ as the \emph{infection rate}. Similarly, an updating node that is infected becomes susceptible with probability $R_{k,m}\, dt$; we refer to $R_{k,m}$ as the \emph{recovery rate}.  (Note that one can also consider functions $F_{k,m}$ and $R_{k,m}$ that depend on network diagnostics other than degree, but we will restrict ourselves to degree.)

It is straightforward to implement these update rules using Monte Carlo simulations. During a small time step $dt$, a node $i$ of degree $k$ with $m$ infected neighbors is endowed with probability $\pi_i=F_{k,m}\, dt$ (if the node is susceptible) or probability $\pi_i = R_{k,m}\, dt$ (if it is infected) of changing its state. By drawing $N$ random numbers $r_i$ from the uniform distribution on $[0,1]$ and comparing the probabilities recorded for each node, the nodes that change states are those nodes $i$ for which $r_i \le \pi_i$. The time step $dt$ should be sufficiently small so that the number of nodes that change state in a single time step is a small fraction of $N$. As discussed in Sec.~\ref{async}, it is common to make $dt$ sufficiently small so that a single node is updated.

We now consider some examples of binary-state dynamics of the type that we just described. We begin with the classical SIS mode of disease spread. We endow this model with a transmission rate of $\lambda$ and a recovery rate of $\mu$. In a small time interval $dt$, each susceptible node has a probability of $\lambda\, dt$ of being infected by each of its infected neighbors. The probability of a susceptible node becoming infected during the time interval $dt$ is then $m\, dt $ as $dt \to 0$ [see equation (\ref{eqn1})]. We thus identify the infection rate for SIS dynamics as
\begin{equation}
	F_{k,m}^{\text{SIS}} = \lambda\, m\,.\label{FSIS}
\end{equation}
Note that equation (\ref{FSIS}) is linear in the number $m$ of infected neighbors. In the SIS model, each infected node can recover to the susceptible state at a constant rate $\mu$. This yields
\begin{equation}
	R_{k,m}^{\text{SIS}} = \mu\,, \label{RSIS}
\end{equation}
which is independent of $m$.

As another example, consider the standard voter model (see Sec.~\ref{votes}). In this example, the labels ``susceptible'' and ``infected'' refer to the two opinions. At each time step, one node is chosen uniformly at random for updating (during the interval $dt=1/N$), and this node copies the state of one of its neighbors (chosen uniformly at random). Therefore, a degree-$k$ node that has $m$ infected neighbors has a probability of $m/k$ of copying an infected neighbor and a probability of $(k-m)/k$ of copying a susceptible neighbor. This yields
\begin{equation}
	F_{k,m}^\text{Voter} = m/k \quad\text{ and }\quad R_{k,m}^\text{Voter}=(k-m)/k\,.\label{FRVoter}
\end{equation}

Within this general framework, one can consider a wide class of well-studied binary-state dynamics. Table~I in Ref.~\cite{gleeson13} lists the infection rates $F_{k,m}$ and recovery rates $R_{k,m}$ for models such as the Bass model for innovation diffusion \cite{Bass69,Dodds04}, the Ising spin model, the majority-vote model \cite{Liggettbook,deOliveira92}, and threshold opinion models. This unified perspective on stochastic binary-state dynamics also allows general derivation of MF approximations and pair approximations from a higher-order approximation scheme. 

%%%%%%%

\subsection{Approximation Methods for General Binary-State Dynamics}

Reference \cite{gleeson13} derived equations for degree-based mean-field (MF) and pair-approximation (PA) theories for general stochastic binary-state dynamics, which are defined in terms of infection rate $F_{k,m}$ and recovery rate $R_{k,m}$. The MF equations are
\begin{equation}
	\frac{d}{dt}\rho_k = -\rho_k \sum_{m=0}^k R_{k,m} B_{k,m}(\omega) + (1-\rho_k) \sum_{m=0}^k F_{k,m}B_{k,m}(\omega)\,, \label{MF}
\end{equation}
where $B_{k,m}(\omega)$ denotes the binomial term $\left(\begin{array}{c} k \\m\end{array}\right) \omega^m(1-\omega)^{k-m}$ and we recall that $\rho_k(t)$ is the fraction of nodes of degree $k$ that are infected at time $t$. We obtain quantity $\omega(t)$ in terms of $\rho_k(t)$ using equation~(\ref{defw}). If a network contains non-empty degree classes from $k=0$ up to some cutoff $k_\text{max}$, then equation (\ref{MF}) consists of a closed system of $k_\text{max}+1$ nonlinear differential equations. (If some of those degree classes are empty, then there will be fewer differential equations.) Standard numerical methods for simulating differential equations enable one to solve these equations efficiently (see Sec.~\ref{numericalmethods}). For example, if we consider the SI model for disease spread, then $F_{k,m}=\lambda \,m$ and $R_{k,m}=0$. Using the binomial sum
\begin{equation}
	\sum_{m=0}^k m \, B_{k,m}(\omega) = k \,\omega
\end{equation}
in equation (\ref{MF}) yields the MF theory that we derived in equation~(\ref{MFmap2}).

One can obtain better accuracy than the MF equations (\ref{MF}) using the PA equations
\begin{align}
\frac{d\rho_k}{d t} &= -\rho_k \sum_{m=0}^k R_{k,m} B_{k,m}(q_k) + (1-\rho_k)\sum_{m=0}^k F_{k,m}B_{k,m}(p_k)\,,\nonumber\\
	\frac{d p_k}{d t} &= \sum_{m=0}^k \left[ p_k - \frac{m}{k}\right]\left[ F_{k,m} B_{k,m}(p_k) - \frac{\rho_k}{1-\rho_k} R_{k,m} B_{k,m}(q_k)\right] + {\beta^s}(1-p_k)-{\gamma^s}p_k\,,\nonumber\\
	\frac{d q_k}{d t} &= \sum_{m=0}^k \left[ q_k - \frac{m}{k}\right]\left[ R_{k,m} B_{k,m}(q_k) - \frac{1-\rho_k}{\rho_k} F_{k,m} B_{k,m}(p_k)\right] + {\beta^i}(1-q_k)-{\gamma^i}q_k \label{PA}
\end{align}
for the variables $\rho_k(t)$, $p_k(t)$, and $q_k(t)$. The rates $\beta^s$, $\gamma^s$, $\beta^i$, and $\gamma^i$ are determined from these variables. For example,
\begin{equation}
	\beta^s= \frac{\sum_k P_k (1-\rho_k)\sum_m (k-m) F_{k,m}B_{k,m}(p_k)}{\sum_k P_k (1-\rho_k)k(1-p_k)}\,
\end{equation}
is the rate at which $SS$ edges become $SI$ edges. See Ref.~\cite{gleeson13} for details. The quantity $p_k(t)$ [respectively, $q_k(t)$] is the probability that a randomly-chosen neighbor of a susceptible (respectively, infected) degree-$k$ node is infected at time $t$.  The system of equations (\ref{PA}) consists of $3k_\text{max}+1$ differential equations, and it typically gives solutions that are more accurate than the MF equations (\ref{MF}).

Equations (\ref{MF}) and (\ref{PA}) reduce to known MF and PA approximations for well-studied dynamical system on networks. For example, consider the SIS disease model. Inserting $F_{k,m}=\lambda \,m$ and $R_{k,m}=\mu$ from equations~(\ref{FSIS}) and (\ref{RSIS}) into these general equations yields the MF equations  that were derived in Ref.~\cite{PastorSatorras01} as well as the PA equations from Refs.~\cite{Eames02,Levin96}. Similarly, using the voter-model rates from equation (\ref{FRVoter}) yields the MF equations of Ref.~\cite{sood05}.  The PA equations that one obtains in this way \cite{Gleeson11} constitute a dynamical system whose dimensionality lies between those of Refs.~\cite{Pugliese09} and \cite{Vazquez08}.

References~\cite{Gleeson11,gleeson13} derived the general MF and PA equations in (\ref{MF}) and (\ref{PA}), respectively, considering a more complicated approximation scheme that involves what are called \emph{approximate master equations (AME)}. In this AME system, one divides nodes into compartments based both on node state and on the number of infected neighbors. One approximates transitions between compartments by global rates to yield a system of $[(k_\text{max}+1)(k_\text{max}+2)]$  closed nonlinear differential equations (See Sec.~III of Ref.~\cite{gleeson13}.) One can then derive PA equations from the AME system by assuming that the number $m$ of infected neighbors of a susceptible (respectively, infected) node is distributed according to the binomial distribution $B_{k,m}(p_k)$ [respectively, $B_{k,m}(q_k)$] and then using this ansatz to derive equations (\ref{PA}) for $p_k(t)$ and $q_k(t)$. This ansatz is exact when it is correct to treat the neighbors of a node as independent of each other. As we discussed in MF assumption (2) of Sec.~\ref{MFdisc}, dynamical correlations imply that this independence-of-neighbors assumption is not true in general, so the AME solutions typically are more accurate than PA solutions. However, the PA requires the solution of only $3 k_\text{max}+1$ equations. The MF equations (\ref{MF}) result from replacing both $p_k$ and $q_k$ by $\omega$. In this situation, one neglects the dependence between the state of the node itself and the states of its neighbors when the node is updated [see MF assumption (1) of Sec.~\ref{MFdisc}]. This yields a system of $k_\text{max}+1$ equations.

%%%%%%

\subsection{Monotonic Dynamics and Response Functions}\label{Monotone0}

For general binary-state dynamics, an AME system like the one discussed in \cite{gleeson13} is high-dimensional and difficult to analyze. However, in the special case of monotonic threshold dynamics (see Sec.~\ref{examples}), the AME system reduces to a coupled set of just two ODEs. This dramatically simplifies analysis. Note that this reduction is exact---i.e., it does not involve any approximation---so the two-dimensional (2D) system is more accurate than naive PA or MF theories (see Fig.~\ref{fig1}).

 %%%%

 \subsubsection{Monotonic Threshold Dynamics}

Monotonic dynamics allow only one-way transitions, so $R_{k,m}\equiv 0$ for all $k$ and $m$. Threshold dynamics occur when the transition rate has the form
\begin{equation}
	F_{\mathbf{k},m} = \left\{ \begin{array}{cc}
	0\,, & \text{ if }m< M_\mathbf{k} \\
	1\,, & \text{ if }m\ge M_\mathbf{k}
\end{array} \right. \,.\label{threshF}
\end{equation}
This reflects deterministic infection (or activation) of a node (once it is chosen for updating) when $m$ equals or exceeds the threshold level $M_\mathbf{k}$. We have introduced the vector $\kv$ to encode two properties of the nodes: their degree $k$ (a scalar) and their type $r$. (More generally, the type could be a vector ${\bf r}$.) Together, these two properties determine the threshold $M_\kv$ for the nodes in a network. For example, all nodes of type 1 might have the same threshold $M_1$, and all nodes of type 2 might have a common threshold $M_2$. Thus, the set of nodes is partitioned into disjoint sets that are labeled by their degree and their type. The type might be some external label for a node (e.g., the dormitory residence of a student in a university Facebook network \cite{Traud2012}), the assignment of a node into some community \cite{Piikk}, or something else. We combine these labels into a two-vector by writing $\kv=(k,r)$ for the $\kv$ class of nodes. We generalize the degree distribution to the joint distribution $P_\kv$, which gives the probability that a node that is chosen uniformly at random has feature vector $\kv$ (i.e., it has degree $k$ and is of type $r$). If the threshold of the nodes are chosen at random using a process that doesn't depend on $k$ (i.e, the thresholds are independent of the node degrees), then one can factor the $P_\kv$ distribution as $P_\kv= P_k P_r$, where $P_k$ is the degree distribution and $P_r$ is the probability that a node is of type $r$.

By considering a large set of discrete types, it is possible to approximate a continuous distribution of thresholds (e.g., a Gaussian distribution) to a desired level of accuracy. Reference~\cite{gleeson13} demonstrated that the AME system reduces exactly to the pair of ODEs
\begin{align}
	\frac{d\rho}{dt} &= h(\phi)-\rho \,, \label{pair1} \\
	\frac{d\phi}{dt} &= g(\phi) - \phi \,,  \label{pair2}
\end{align}
where
\begin{align}
%\label{4}
	h(\phi) &= \rho(0) + [1-\rho(0)]\sum_\kv P_\kv  \sum_{m\ge M_\kv} B_{k,m}(\phi) \,, \label{4} \\
%\end{align}
%and
%\begin{align}\label{5}	
	g(\phi) &= \rho(0)+[1-\rho(0)]\sum_\kv \frac{k}{z} P_\kv \sum_{m\ge M_\kv} B_{k-1,m}(\phi) \label{5} \,.
\end{align}
The variable $\phi(t)$ is the probability that a node attached to one end of an edge chosen uniformly at random is infected, conditional on the node attached to the other end of the edge being susceptible \cite{gleeson08}.
The initial conditions for equations (\ref{pair1},\ref{pair2}) are
\begin{equation}\label{ic}
	 \phi(0)=\rho(0) = \sum_\kv P_\kv \rho_\kv(0)\,.
\end{equation}	
Solving (\ref{pair1},\ref{pair2}) with the initial conditions (\ref{ic}) yields the fraction $\rho(t)$ of infected nodes in a network at time $t$ to a very high level of accuracy. See Fig.~\ref{fig1} for some examples. Note that the AME solutions (red curves) are identical to the solution of the 2D system (\ref{pair1},\ref{pair2}), and they match the numerical simulation results (black symbols) very closely.

%%%%%%%

\subsubsection{Response Functions for Monotonic Binary Dynamics}\label{deriveMFPA}

Although Ref.~\cite{gleeson13} derived equations (\ref{pair1},\ref{pair2}) from a full AME system, they can also be obtained using other approaches. Reference~\cite{gleeson08}, for example, derived these equations as the asynchronous limit of corresponding synchronous-updating equations that were determined by using methods from the study of zero-temperature random-field Ising models \cite{Dhar97,gleeson07}. These methods are related to belief-propagation and message-passing algorithms \cite{Mezardbook}, and the approach of Ref.~\cite{gleeson08} makes it possible to generalize beyond the (uncorrelated) configuration-model networks that we have considered thus far. Low-dimensional descriptions such as (\ref{pair1},\ref{pair2}) for monotonic threshold dynamics have also been derived for networks with degree-degree correlations and/or community structure \cite{gleeson08,Piikk}, networks with non-negligible clustering \cite{Adamcascades}, and multiplex networks \cite{Yagan}. In Ref.~\cite{gleeson08}, it was also demonstrated that one can express several structural characteristics of networks, such as the sizes of $K$-cores \cite{Dorogovtsev06} and the giant connected component (GCC) sizes for site and bond percolation, as steady states of monotonic threshold dynamics.  Such characteristics can thereby be determined using equations of the form (\ref{pair1},\ref{pair2}). This perspective is likely to be particularly fruitful in the extension of traditional network measures to multiplex and other multilayer networks \cite{Cellai13,Yagan,mikko-review,bocca-review}.

As an example, we show how to apply equations (\ref{pair1},\ref{pair2}) to the Watts threshold model for social influence (which we described in Sec.~\ref{thresh}). Recall that each node has a threshold $R_i$ that is drawn from some distribution $P_R$, and a node $i$ that is being updated becomes active (i.e., moves to the ``infected'' state) if its fraction of active neighbors $m_i/k_i \geq R_i$.  The thresholds $R_i$ are distributed independently of node degrees, so $P_\kv=P_kP_R$. Equation~(\ref{threshF}) for this model is
\begin{equation}
	F_{\mathbf{k},m} = \left\{ \begin{array}{cc}
		0\,, & \text{ if }m< k R \\
		1\,, & \text{ if }m\ge k R
	\end{array} \right. \,,\label{specificthreshF}
\end{equation}
and the sums over types in equation (\ref{4}) become
\begin{align}\label{here}
	\sum_{\kv} P_\kv \sum_{m\ge M_\kv} B_{k,m}(\phi) & = \sum_R P_R \sum_{m\ge k R }B_{k,m}(\phi) \\
	& = \sum_{m=0}^k B_{k,m}(\phi) \int_{-\infty}^\infty P_R H\left[m-k R\right]\, dR \\
	& = \sum_{m=0}^k B_{k,m}(\phi) \int_{-\infty}^{\frac{m}{k}} P_R \, dR \\
	& = \sum_{m=0}^k B_{k,m}(\phi) C\left(\frac{m}{k}\right)\,,
\end{align}
where $H$ denotes the Heaviside function and $C$ is the cumulative distribution function (CDF) of the thresholds.

Reference \cite{gleeson08} showed that the $C(m/k)$ term in equation (\ref{here}) is an example of a \emph{response function} for monotonic dynamics. A response function encapsulates the mechanism by which a susceptible node becomes infected when it is updated. For the present discussion, we suppose that a response function $f(k,m)$ depends on the degree $k$ of a node and the number $m$ of infected neighbors. (More generally, a response function can also depend on other network characteristics.) In this setting, equations~(\ref{4},\ref{5}) take the form
\begin{align}
	h(\phi) &= \rho(0) + [1-\rho(0)]\sum_k P_k  \sum_{m=0}^k B_{k,m}(\phi) f(k,m)\,, \label{hgen} \\
	g(\phi) &= \rho(0)+[1-\rho(0)]\sum_k \frac{k}{z} P_k \sum_{m=0}^{k-1} B_{k-1,m}(\phi) f(k,m) \,. \label{ggen}
\end{align}
Response functions of this type have also been defined for other monotonic threshold models (such as many of the complex-contagion models in Sec.~\ref{examples}) and related problems. In bond percolation, for example, the equations for the fractional size of the GCC (i.e., the epidemic size in an SIR model), which was previously derived using generating-function methods \cite{Callaway00}, are reproduced by using $\rho(0)=0$, considering the response function
\begin{equation}
	f(k,m) = \left\{ \begin{array}{cc}
		0\,, & \text{ if }m=0 \\
		1-(1-p)^m\,, & \text{ if }m>0
	\end{array} \right.\,,\label{percf}
\end{equation}
and applying the identity
\begin{equation}
	\sum_{m=1}^k B_{k,m}(\phi)\left[1-(1-p)^m\right] = 1-(1-p\phi)^k\,.
\end{equation}
See Sec.~II.B of \cite{gleeson08} for additional discussion.

%%%%%%%

\subsubsection{Cascade Conditions} \label{subsubcascade}

In addition to giving very accurate predictions for monotonic dynamics, equations (\ref{pair1},\ref{pair2}) make it possible to obtain analytical insights into dynamical processes.  Let's again take the Watts threshold model as an example.  Consider the question of whether, for a given distribution of thresholds, global cascades can occur on configuration-model networks with degree distribution $P_k$. Watts used percolation theory in \cite{watts2002} to derive a \emph{cascade condition} that addresses this question.  We will now derive the same cascade condition by using linear stability analysis of the 2D dynamical system defined by equations (\ref{pair1},\ref{pair2},\ref{hgen},\ref{ggen}).

Suppose that all nodes have positive thresholds, so $C(0)=0$, and consider the seed fraction $\rho(0)$ to be vanishingly small. The system (\ref{pair1},\ref{pair2}) then has an equilibrium point at $(\rho,\phi) = (0,0)$ that corresponds to a complete absence of infection. However, if this equilibrium point is (linearly) unstable, then a small perturbation (e.g., the infection of a single node) can move the dynamical system away from the equilibrium at $(0,0)$ and result in values $\rho(t) > 0$, constituting a global cascade.  By the term ``global'', we mean that a non-vanishing fraction of the nodes in the network become infected in the $N\to\infty$ limit of infinite network size.

The stability of system (\ref{pair1},\ref{pair2}) is controlled by the second equation (\ref{pair2}). When this 1D dynamical system is unstable at $\phi=0$, the values of $\rho$ determined from equation (\ref{pair1}) are also strictly positive. One uses standard linear stability analysis \cite{Drazinbook,stro1994} to show that (local) instability occurs at $\phi=0$ if and only if the following condition holds:
\begin{equation}
	\left.\frac{d}{d\phi}\left[g(\phi)-\phi\right]\right|_{\phi=0}>0\,.
\end{equation}
That is, there is a local instability when $g'(0)>1$, and the monotonic nature of the dynamics guarantees in this case that $\rho(t)$ is strictly positive for all time (i.e., that a global cascade occurs). We now differentiate equation (\ref{ggen}), which is the general response-function form of $g$, using $f(k,m)=C(m/k)$ for the Watts model and incorporating the facts that $B_{k,m}(0)=\delta_{m,0}$ and $f(k,0)=0$. This yields the cascade condition that was found in \cite{watts2002}:
\begin{equation}\label{casccond}
	\sum_k \frac{k}{z}(k-1) P_k  f(k,1) > 1\,.
\end{equation}
When equation (\ref{casccond}) is satisfied, global cascades can occur in the Watts model. Note that this general condition incorporates information about both network topology (via the degree distribution $P_k$) and node-level dynamics (through the expected response $f(k,1)$ of a degree-$k$ node to a single infected neighbor). It thereby yields interesting and useful information on how network topology influences dynamics.

In closing, we note that there are several ways to define a ``cascade'', especially in the context of applications.  When studying purely empirical data, one might wish to define a ``cascade'' based on a specified minimum fraction of nodes that eventually become infected or a specified minimum fraction that become infected within a specified amount of time.  In practical situations, it can be important to consider the ``infection'' of people with a meme within a finite duration of time, so the $t \rightarrow \infty$ limit that one typically considers to compute cascade sizes is too restrictive on some occasions. (For example, looser notions of a cascade are relevant to consideration of social influence on networks in the commercial and governmental sector, as it may be necessary to convince as many people as possible to adopt an idea in a very limited amount of time.)

%%%%%%

\end{document}